\documentstyle[12pt,aps,psfig]{revtex}
\begin{document}

\def\be{\begin{equation}}
\def\ee{\end{equation}}
\def\ba{\begin{eqnarray}}
\def\ea{\end{eqnarray}}                                                        

\title{Rate Equations for Sympathetic Cooling of Trapped Bosons or
  Fermions} 
\author{T. Papenbrock$^1$, A. N. Salgueiro$^2$, and
  H. A. Weidenm\"uller$^2$}
\address{$^1$ Physics Division, Oak Ridge National Laboratory, Oak
  Ridge, TN 37831, USA}
\address{$^2$Max-Planck-Institut f\"ur Kernphysik, D-69029 Heidelberg,
  Germany}

\date{\today}

\maketitle

\begin{abstract}

We derive two different sets of rate equations for sympathetic cooling
of harmonically trapped Bosons or Fermions. The rate equations are
obtained from a master equation derived earlier by Lewenstein {\it et
al.} [Phys. Rev. A {\bf 51} (1995) 4617] by means of decoherence and
ergodicity arguments. We show analytically that the thermal
equilibrium state is a stationary solution of our rate equation. We
present analytical results for the rate coefficients which are needed
to solve the rate equations, and we give approximate formulae that
permit their computation in practice.  We solve the two sets of rate
equations numerically and compare the results. The cooling times
obtained in both approaches agree very well.  The equilibration rates
show fair agreement.

\end{abstract}

\section{Introduction}
\label{int}

The cooling of atoms in traps is an important tool in the study of the
behavior of systems of Bosons and Fermions at low temperatures.
Usually, the last step in the cooling process is evaporative cooling.
For this step to be efficient, thermodynamic equilibrium of the cooled
gas must be (nearly) attained at all times. This condition is met if
the atoms interact sufficiently strongly. There is a number of gases
in which the interaction is too weak for evaporative cooling to work.
In such cases, one resorts to ``sympathetic cooling'': Another gas is
cooled evaporatively. This gas acts as the cooling agent for the gas
which is to be cooled. It is usually legitimate to assume that the
number of atoms in the cooling agent is very large, or that another
mechanism is operative which keeps the cooling agent at fixed
temperature. In either case, the cooling agent acts as a heat bath of
fixed temperature.

The method of sympathetic cooling was proposed almost twenty years
ago\cite{win78,dru80} and has since found widespread application (see
Refs.~\cite{lar86,van85}). Exciting recent applications include the
sympathetic cooling of ${}^6$Li Fermions in a bath of ${}^7$Li Bosons
\cite{truscott,schreck}, and the production of dual Bose--Einstein
condensates with sympathetic cooling\cite{myatt,delannoy}.  However,
to the best of our knowledge, there does not exist until now a
practicable theoretical description of this process. This gap is
remarkable because theory would be expected to make predictions on the
dependence of the cooling rate on various parameters defining both the
interaction between atoms in the cooling agent and in the cooled gas,
and on the trap potential, and might thus be helpful in improving the
cooling process. We are aware of only two papers which deal with
sympathetic cooling theoretically.  Lewenstein et al.~\cite{lew95}
derived a master equation for sympathetic cooling. Unfortunately, that
master equation is too complex to be useful, see below. Geist {\it et
al.}~\cite{Geist} have formulated sympathetic cooling of Fermi gases
in terms of a quantum Boltzmann equation. The tremendous
simplification of the collision matrix elements achieved in this way
results, however, mainly in a qualitative description of the cooling
process.

It is the purpose of the present work to derive rate equations for the
cooling process which can be implemented and used practically. We do
so by simplifying the master equation of Ref.~\cite{lew95} and by
reducing it to a set of rate equations. We demonstrate the utility of
our approach by presenting analytical and numerical results which
apply to the cooling process. In the present Section, we begin by
listing the assumptions and collecting the results of
Ref.~\cite{lew95}. We use the notation of Ref.~\cite{lew95}.

The cooling gas (referred to as system $B$) consists of Bosons in
thermal equilibrium at temperature $T_B$. We use $\beta_B = (\kappa
T_B)^{-1}$ where $\kappa$ is Boltzmann's constant. It is assumed that
$T$ does not change during the cooling process. It is also assumed
that the atoms in system $B$ are much heavier than those of the cooled
gas. The single--particle states of system $B$ are then approximated
by plane waves with energies $\varepsilon(k) = (\hbar {\vec k})^2 / (2
M)$ with $\vec k$ the momentum and $k = |{\vec   k}|$, with
corresponding creation and annihilation operators $b^{\dagger}({\vec
k})$ and $b({\vec k})$, respectively. Summations over $\vec k$ are
always replaced by integrations. The number $n(k)$ of particles with
wave number $\vec k$ is given by 
\begin{equation}
\label{eq1}
n(k) = \frac{z \exp[-\beta_B \varepsilon(k)]}{1 - z \exp[-\beta_B
  \varepsilon(k)]}
\end{equation}
where $z$ denotes the fugacity. While we retain Eq.~(\ref{eq1}) in our
analytical work, in the numerical work of Section~\ref{num} below
$T_B$ is taken to be so large that the coefficients $n(k)$ can be
replaced by a Boltzmann distribution, 
\begin{equation}
\label{eq2}
n(k) = n_B \Lambda_B^3 \exp[-\beta_B \varepsilon(k)] \ ,
\end{equation}
with 
\begin{equation}
\label{eq2a}
n_B = (2 \pi)^{-3} \int {\rm d}^3 k \ n(k) 
\end{equation}
the density of atoms in $B$ and $\Lambda_B  = (2 \pi \hbar^2 \beta_B
/ M)^{1/2}$ the thermal de--Broglie wave length of the atoms in system
$B$. Again, Eq.~(\ref{eq2}) applies if the mass of the atoms in
system $B$ is sufficiently large.

The atoms subject to sympathetic cooling (referred to as system $A$)
are confined by a harmonic--oscillator potential with trap frequency
$\nu$. We simplify the notation by confining ourselves to isotropic
traps in three dimensions. The generalization to other dimensions, and
to non--isotropic traps, is straightforward. The single--particle
states in the harmonic--oscillator potential have quantum numbers
${\vec n} = n_x, n_y, n_z$. The integers $n_x, n_y, n_z$ are positive
or zero and count the number of harmonic--oscillator quanta in the
$x,y$ and $z$ directions of a Cartesian coordinate system. The
single--particle eigenfunctions are labelled $\psi_{\vec n}({\vec
  x})$, the eigenvalues $\varepsilon_{\vec n}$. The corresponding
creation and annihilation operators are denoted by $a^{\dagger}_{\vec
  n}$ and $a_{\vec n}$, respectively. Unless stated otherwise, we
consider the atoms in system $A$ to be Bosons.

The interaction $H_{A-B}$ between atoms in system $A$ and those in
system $B$ is described by the Hamiltonian
\begin{equation} 
\label{3a}
H_{A-B} = \sum_{\vec n,{\vec n}'} \int d\vec k d{\vec k}' \gamma_{\vec
  n,{\vec n}'}(\vec k, {\vec k'})a_{\vec n}^{\dagger}a_{{\vec n}'}
b^{\dagger}(\vec k) b({\vec k}') \ . 
\end{equation}
The matrix elements of the interaction are denoted by $\gamma_{{\vec
  n}, {\vec n}'}({\vec k},{\vec k}')$. These matrix elements contain
the two--body interaction which is approximated by a delta function of
strength
\begin{equation}
\label{eq3}
C = \frac{4 \pi \hbar^2 a}{2 \mu}
\end{equation}
where $a$ is the scattering length and $\mu$ the reduced mass. Then,
the matrix elements $\gamma_{{\vec n}, {\vec n}'}({\vec k},{\vec k}')$
are given by
\begin{equation}
\label{eq4}
\gamma_{{\vec n},{\vec n}'}({\vec k},{\vec k}') = \frac{C}{(2 \pi)^3}
\int {\rm d}^3 x \ \psi_{\vec n}({\vec x})^{*} \psi_{{\vec n}'}({\vec
  x}) \exp{\left[ -i \left({\vec k} - {\vec k}'\right) {\vec x}\right]} \ .
\end{equation}
The rate coefficients $\Gamma^{{\vec m},{\vec m}'}_{{\vec n},{\vec
    n}'}$ appearing in the master equation below are given by
\begin{eqnarray}
\label{eq5}
&&\Gamma^{{\vec m},{\vec m}'}_{{\vec n},{\vec n}'} = \frac{1}{2
  \hbar^2} \int_{-\infty}^{\infty} {\rm d} \tau \int {\rm d}^3 k {\rm
  d}^3 k' \gamma_{{\vec n},{\vec n}'}({\vec k},{\vec k}')
  \gamma_{{\vec m},{\vec m}'}({\vec k}',{\vec k}) \nonumber \\
&&\qquad \qquad \times n(k) [n(k')+1] \exp{\left[ i\left(\varepsilon(k) -
  \varepsilon(k') + \alpha \hbar \nu\right) \tau / \hbar \right]} \ .
\end{eqnarray}
The integer $\alpha$ is defined by
\begin{equation}
\label{eq6}
\alpha = (m'_x+m'_y+m'_z) - (m_x+m_y+m_z) = (n_x+n_y+n_z) - (n'_x+n'_y+n'_z) 
\ .
\end{equation}
Obviously, $\alpha$ can be positive, negative, or zero. The last
relation in Eq.~(\ref{eq6}) restricts the possible values of ${\vec
  m}, {\vec m}', {\vec n}, {\vec n}'$.

The master equation describes the dependence on time $t$ of the
reduced density matrix $\rho_A(t)$ for system $A$, obtained by tracing
the total density matrix for systems $A$ plus $B$ over system $B$.
Under the assumptions that the process is Markovian and that the
correlation time for the interaction between systems $A$ and $B$ is
much shorter than the cooling time, and with the help of a
rotating--wave approximation, the master equation takes the form 
\begin{equation}
\label{eq7}
\frac{{\rm d} \rho_A(t)}{{\rm d}t} = - \frac{i}{\hbar} \biggl [ H_A +
H'_{A-A}, \rho(t)_A \biggr ] + {\cal L} \rho_A \ . 
\end{equation}
Here, $H_A$ is the sum of the single--particle Hamiltonians for atoms
in system $A$ containing the kinetic energy and the
harmonic--oscillator potential of the trap, while $H'_{A-A}$ contains
the interaction between the atoms in system $A$. The action of the
Liouvillean $\cal L$ on the reduced density matrix $\rho_A(t)$ is
given by
\begin{eqnarray}
\label{eq8}
&&{\cal L} \rho_A = \sum_{{\vec n}, {\vec n}', {\vec m}, {\vec m}'}
\Gamma^{{\vec m},{\vec m}'}_{{\vec n},{\vec n}'} \biggl ( 2
a^{\dagger}_{\vec m} a_{{\vec m}'} \rho_A(t) a^{\dagger}_{\vec n}
a_{{\vec n}'} - a^{\dagger}_{\vec n} a_{{\vec n}'} a^{\dagger}_{\vec
  m} a_{{\vec m}'} \rho_A(t) \nonumber \\
&&\qquad \qquad \qquad \qquad \qquad - \rho_A(t) a^{\dagger}_{\vec n}
a_{{\vec n}'} a^{\dagger}_{\vec m} a_{{\vec m}'} \biggr ) \ .
\end{eqnarray}

Eqs.~(\ref{eq7},\ref{eq8}) are not useful as they stand. Even if the
accessible single--particle states of the harmonic oscillator are
restricted to the lowest $30 \hbar \nu$ or so, the dimension of the
resulting matrix representation of $\rho_A$ is enormous. Moreover, the
numerical calculation of each of the rate coefficients $\Gamma^{{\vec
    m},{\vec m}'}_{{\vec n},{\vec n}'}$ presents formidable
difficulties. These are compounded by the sheer number of such
coefficients needed in the calculation. Thus, it is necessary to
simplify Eqs.~(\ref{eq7},\ref{eq8}). We do so in a sequence of steps.
In Section~\ref{dec}, we show that the decoherence time $\tau_{\rm
  dec}$ for the system $A$ is inversely proportional to $\sqrt{n_B}$
where $n_B$ is proportional to the number of particles in system $B$,
see Eq.~(\ref{eq2a}). Therefore, $\tau_{\rm dec}$ is the smallest time
scale in the problem. As a consequence, the off--diagonal elements of
$\rho_A$ are damped out very quickly, and $\rho_A$ becomes diagonal in
energy representation. The ensuing reduction of the number of
equations and of rate coefficients is not yet sufficient, however, to
lead to a practicable problem. We aim at a sufficiently simple rate
equation. In Ref.~\cite{lew95} it was pointed out that the degeneracy
of the harmonic--oscillator states causes difficulties in the
conversion of the master equation into a rate equation. We overcome
this problem by assuming that the interaction between atoms in system
$A$, although weak, lifts this degeneracy. We investigate two
independent possibilities of reducing the problem further. In
Section~\ref{mic}, we average the density matrix over all many--body
states having, for $H_{A-A}' = 0$, the same excitation energy (a fixed
multiple of $\hbar \nu$). The resulting rate equation connects
diagonal matrix elements of $\rho_A$ in energy representation.
Alternatively, we introduce a factorization assumption for the density
matrix $\rho_A$ (Section~\ref{fac}). This procedure yields a simple
and intuitively convincing form of the rate equation. The
factorization assumptions used in Section~\ref{fac} and, more
implicitly, also in Section~\ref{mic} are kin to a mean--field
approximation. In Section~\ref{fac} we check their validity by
calculating the influence of the fluctuations on the rate equation
which are neglected in the mean--field approach. In Section~\ref{con},
we test the consistency of our scheme. We show that the equilibrium
distribution of Bosons in the ground state of system $A$ derived from
our rate equation coincides with the one calculated by
Scully~\cite{scu99} from a different starting point. We then use our
results for a numerical calculation. The rate coefficients are worked
out in Section~\ref{rat}. The rate equations are solved, and the
results are discussed in Section~\ref{num}. Particular attention is
paid to a comparison between results obtained in the framework of
Section~\ref{mic} and of Section~\ref{fac}. Section~\ref{conc}
contains the conclusions.

\section{Decoherence}
\label{dec}

The interaction with the heat bath (i.e., with system $B$) not only
cools system $A$ but also induces decoherence: The off--diagonal
elements (in energy representation) of the reduced density matrix 
$\rho_A(t)$ decay in time. In this Section, we estimate the
decoherence time $\tau_{\rm dec}$ and show that $\tau_{\rm dec}$ is
proportional to $1 / \sqrt{n_B}$ and, thus, much shorter than all
other time scales in the problem.

The common procedure~\cite{dec} consists in calculating the time
dependence of the linear entropy $\delta_A(t)$ defined by
\begin{equation}
\label{eq9}
\delta_A(t) = 1 - {\rm tr}_A ([\rho_A(t)]^2)
\end{equation}
in terms of a power--series expansion in $t$, assuming that at time
$t=0$ the total density matrix $\rho(0)$ of systems $A+B$ obeys
$\rho(0) = \rho_A(0) \otimes \rho_B(0)$ with ${\rm tr}_A \rho_A(0) =
{\rm tr}_A (\rho_A(0))^2 = 1$ and ${\rm tr}_B \rho_B(0) =1$. The
decoherence time $\tau_{\rm dec}$ is given by the inverse of the
coefficient multiplying $t$ or, should this term vanish, by the square
root of the inverse of the coefficient multiplying $t^2$. We face the
second alternative because in Ref.~\cite{lew95} it is assumed that
${\rm tr}_B (\rho_B H_{A-B}) = 0$. This causes the term linear in $t$
to vanish.

Second--order perturbation theory with respect to $H_{A-B}$ yields
\begin{eqnarray}
\label{eq10}
\delta_A(t) &=& \frac{2}{\hbar^2} \int_0^t {\rm d}t' \int_0^{t'} {\rm
  d}t'' \nonumber \\
&&\times {\rm tr}_A \biggl ( \rho_A(0) {\rm tr}_B \left( \left[
  {\tilde H}_{A-B}(t') , \left[{\tilde H}_{A-B}(t'') , \rho_A(0)
  \otimes \rho_B(0) \right] \ \right] \ \right) \biggr ) . 
\end{eqnarray}
The tilde indicates that the operator is taken in the interaction
representation. We focus attention on the trace over $B$ and consider
\begin{equation}
\label{eq11} 
X = {\rm tr}_B \left( \left[ {\tilde H}_{A-B}(t') , \left[{\tilde
  H}_{A-B}(t''), \rho_A(0) \otimes \rho_B(0) \right] \ \right] \
  \right) \ .
\end{equation}
With $H_0 = H_A + H_B$ the sum of the Hamiltonians for the atoms in
systems $A$ and $B$, $X$ is explicitly given by
\begin{equation}
\label{eq12} 
X = \sum_{{\vec n}_1 {\vec n}'_1} \sum_{{\vec n}_2 {\vec n}'_2} \int
{\rm d}^3 k_1 \int {\rm d}^3 k'_1 \int {\rm d}^3 k_2 \int {\rm d}^3
k'_2 \gamma_{{\vec n}_1 {\vec n}'_1}({\vec k}_1, {\vec k}'_1)
\gamma_{{\vec n}_2 {\vec n}'_2}({\vec k}_2, {\vec k}'_2) \ Y  
\end{equation}
where
\begin{eqnarray}
\label{eq13}
Y &=& {\rm tr}_B \biggl ( \Bigl[ \exp(i H_0 t' / \hbar)
    a^{\dagger}_{{\vec n}_1} a_{{\vec n}'_1} b^{\dagger}_{{\vec k}_1}
    b_{{\vec k}'_1} \exp(-i H_0 t' / \hbar) , \nonumber \\
&& \left[ \exp(i H_0 t'' / \hbar) a^{\dagger}_{{\vec n}_2} a_{{\vec
    n}'_2} b^{\dagger}_{{\vec k}_2} b_{{\vec k}'_2} \exp(-i H_0 t'' /
    \hbar) , \rho_A(0) \otimes \rho_B(0) \right] \ \Bigr] \ \biggr ) \
    .
\end{eqnarray}
We define
\begin{equation}
\label{eq14}
{\cal A}_{{\vec n}_1 {\vec n}'_1}(t) = \exp(i H_A t / \hbar)
a^{\dagger}_{{\vec n}_1} a_{{\vec n}'_1} \exp(-i H_A t / \hbar)
\end{equation}
and correspondingly for ${\cal B}_{{\vec k}_1 {\vec k}'_1}(t)$. A
straightforward calculation yields
\begin{eqnarray}
\label{eq15}
Y &=& \left[ {\cal A}_{{\vec n}_1 {\vec n}'_1}(t') , {\cal A}_{{\vec
  n}_2 {\vec n}'_2}(t'') \rho_A(0) \right] \ {\rm tr}_B \left( {\cal
  B}_{{\vec k}_1 {\vec k}'_1}(t') {\cal B}_{{\vec k}_2 {\vec
  k}'_2}(t'') \rho_B(0) \right) \nonumber \\
&& - \left[ {\cal A}_{{\vec n}_1 {\vec n}'_1}(t'),  \rho_A(0) {\cal
  A}_{{\vec n}_2 {\vec n}'_2}(t'') \right] \ {\rm tr}_B \left( {\cal
  B}_{{\vec k}_1 {\vec k}'_1}(t') \rho_B(0) {\cal B}_{{\vec k}_2 {\vec
  k}'_2}(t'') \right) \ .
\end{eqnarray}
The two traces over $B$ yield
\begin{eqnarray}
\label{eq16}
{\rm tr}_B ( {\cal B}_{{\vec k}_1 {\vec k}'_1}(t') {\cal B}_{{\vec
  k}_2 {\vec k}'_2}(t'') \rho_B(0) ) &=& n(k_1) [ n(k_2) + 1]
  \nonumber \\
&& \times \exp \Bigl(i \left(\varepsilon(k_1) -
  \varepsilon(k_2)\right)\left(t' - t''\right)/ \hbar \Bigr) \ ,
  \nonumber \\
{\rm tr}_B ( {\cal B}_{{\vec k}_1 {\vec k}'_1}(t') \rho_B(0) {\cal
  B}_{{\vec k}_2 {\vec k}'_2}(t'') ) &=& n(k_2) [ n(k_1) + 1 ]
  \nonumber \\
&& \times \exp \Bigl(i \left(\varepsilon(k_1) -
  \varepsilon(k_2)\right)\left(t' - t''\right) / \hbar \Bigr) \ .
\end{eqnarray}
Inserting this result back into Eq.~(\ref{eq15}) and the latter into
Eq.~(\ref{eq12}), performing the trace over $A$, carrying out the time
integrations, and expanding in powers of $t$, we obtain to lowest
non--vanishing order
\begin{equation}
\label{eq17}
\delta_A(t) = \biggl ( \frac{t}{\tau_{dec}} \biggr )^2
\end{equation}
where
\begin{eqnarray}
\label{eq18}
\biggl ( \frac{1}{\tau_{dec}} \biggr )^2 &=& \frac{1}{\hbar^2}
  \sum_{{\vec n}_1 \neq {\vec n}_2} \int {\rm d}^3 k_1 \int {\rm d}^3
  k_2 |\gamma_{{\vec n}_1 {\vec n}_2}({\vec k}_1, {\vec k}_2)|^2
  \nonumber \\
&& \times \biggl ( n(k_1) [ n(k_2) + 1] n({\vec n}_1) [ n({\vec n}_2)
  + 1] \nonumber \\
&& + n(k_2) [ n(k_1) + 1] n({\vec n}_2) [ n({\vec n}_1) + 1] \biggr ) \ . 
\end{eqnarray}
Here, $n({\vec n}) = {\rm tr}_A ( a^{\dagger}_{\vec n} a_{\vec n}
\rho_A(0))$.

The terms containing the occupation numbers $n(k)$ in Eq.~(\ref{eq18})
have the form $n(k_1) [ n(k_2) + 1] = n(k_1) n(k_2) + n(k_1)$ and
$n(k_2) [ n(k_1) + 1] = n(k_1) n(k_2) + n(k_2)$. We focus attention on
the terms linear in the occupation numbers and observe that the
$k$--dependence of the remaining part of the integrand resides in 
$|\gamma_{{\vec n}_1 {\vec n}_2}({\vec k}_1, {\vec k}_2)|^2$. This
quantity depends only on ${\vec k}_1 - {\vec k}_2$, see Eq.~(\ref{eq4}).
In the term linear in $n(k_1)$ ($n(k_2))$, we introduce the integration
variables ${\vec k}_1$ and ${\vec \kappa} = {\vec k}_1 - {\vec k}_2$
(${\vec k}_2$ and ${\vec \kappa} = {\vec k}_1 - {\vec k}_2$,
respectively). The integration over ${\vec k}_1$ (${\vec k}_2$,
respectively) then yields $n_B$, see Eq.~(\ref{eq2a}). This shows that
$\tau_{\rm dec} \sim 1 / \sqrt{n_B}$.

In order to implement decoherence, we follow Ref.~\cite{lew95} and
assume that the interaction $H'_{A-A}$ between atoms in system $A$ is
weak. Indeed, sympathetic cooling is used precisely when this
condition is met and other procedures fail. To quantify this
assumption, we observe that in the absence of $H'_{A-A}$, a suitable
basis for $\rho_A(t)$ is given by the many--body eigenstates $|M j
\rangle$ of the trap, i.e., of the three--dimensional harmonic
oscillator. Here $M$ denotes the total excitation energy $M \ \hbar
\nu$. The running index $j$ labels the degenerate states. As we drop
the condition $H'_{A-A} = 0$, the degeneracy of the states $|M j
\rangle$ is lifted, and even states belonging to different values $M$
get mixed. We quantify the condition that the atoms in system $A$
interact weakly by the requirement that the latter mixing is
negligible. We diagonalize $H_A + H'_{A-A}$ and label the resulting
eigenstates by $| M \mu \rangle$, the eigenvalues by $\varepsilon(M,
\mu)$. Here, $M$ has the same meaning as before, while $\mu$ is a new
running label.

It turns out that for $H'_{A-A}$ weak, the master equation has a
curious feature. To see this, we take the matrix element of
Eqs.~(\ref{eq7},\ref{eq8}) between a state $\langle M_1 \mu_1 |$ and a
state $| M_2 \mu_2 \rangle$. Inspection shows that the master equation
connects the time derivative of $\langle M_1 \mu_1 | \rho_A(t) |
M_2 \mu_2 \rangle$ with the values of $\langle (M_1 + \alpha)
\mu_3 | \rho_A(t) | (M_2 + \alpha) \mu_4 \rangle$ with
$\alpha$ defined in Eq.~(\ref{eq6}). In other words, the master
equation takes the form of sets of coupled equations, the equations in
each set characterized by a fixed value of $M_1 - M_2$. The equations
in different sets are not coupled with each other. In this situation,
decoherence shows that the matrix elements $\langle M_1 \mu_1 |
\rho_A(t) | M_2 \mu_2 \rangle$ with $M_1 \neq M_2$ vanish rapidly,
and that it suffices to consider only a single set, containing the
diagonal block $\langle M \mu_1 | \rho_A(t) | M \mu_2 \rangle$.
Within this block, we may again use decoherence to argue that of these
matrix elements, only the ones with $\mu_1 = \mu_2$ survive. This is
what we assume from now on.

The ensuing reduction in the number of equations is, of course,
enormous: Let the total number of states $| M j \rangle$ be $N$. Then,
the number of equations needed to determine $\rho_A(t)$ is $N^2$ while
the number remaining after we have used the decoherence argument is
just $N$. However, this number is still too large, and a further
simplification of the master equation is needed.
 
\section{Microcanonical Average}
\label{mic}

Because of decoherence, the reduced density matrix $\rho_A$ is
diagonal in the basis of states $|M \mu \rangle$. We now assume, in
addition, that for fixed $M$, all matrix elements of $\rho_A$ are
equal. This ergodicity-assumption seems physically reasonable: We
expect the equilibration time of the entire system $A$ to be much
larger than the time it takes to equilibrate states which have
(nearly) the same energy. Thus,
\begin{equation}
\label{eq19}
\langle M_1 \mu_1 | \rho_A(t) | M_2 \mu_2 \rangle = 
{p_M(t)\over d(M,N_A)} \, \delta_{M_1,M_2}
\delta_{\mu_1,\mu_2} \ .
\end{equation}
The Kronecker delta's express decoherence, and $p_M/d(M,N_A)$ is the
common value. We define $d(M,N_A)$ as the number of levels belonging
to fixed $M$ for a system having $N_A$ particles. Transforming back
to the basis of states $|M j \rangle$, we see immediately that for
fixed $M$, $\rho_A(t)$ is diagonal in that basis, too, with diagonal
elements $p_M(t)$. We introduce an explicit notation for the states
$|M j \rangle$ in terms of the occupation numbers $n_{\vec m}$ of the
single--particle states of the harmonic oscillator. These are subject
to the constraints $\sum_{\vec m} n_{\vec m} = N_A$ where $N_A$ is the
number of atoms in system $A$, and $\sum_{\vec m} n_{\vec m}
\varepsilon_{\vec m} = M \hbar\nu$. The states $|M j \rangle$ are
written explicitly as $|M \lbrace n_{\vec m} \rbrace \rangle$.  We
observe that the $p_M$ introduced in Eq.~(\ref{eq19}) obey the sum
rule $\sum_M p_M = 1$.

Taking the partial trace of the master equation over the states $|M
\lbrace n_{\vec m} \rbrace \rangle$ with fixed $M$ and using
Eq.~(\ref{eq19}), we obtain
\begin{eqnarray}
\label{eq20}
\frac{{\rm d} p_M}{{\rm d}t} &=& 2 \sum_{{\vec m}{\vec
      m}'{\vec n}{\vec n}'} \Gamma^{{\vec m},{\vec m}'}_{{\vec
      n},{\vec n}'} \times \nonumber \\
&&\Biggl ( \sum_{M'} {p_{M'}\over d(M',N_A)} 
      \sum_{{\vec \lambda}{\vec \kappa}} 
           \langle M' \{ n_{\vec \kappa}
      \} | a^{\dagger}_{\vec n} a_{{\vec n}'} | M \{ n_{\vec \lambda}
      \} \rangle
      \langle
      M \{ n_{\vec \lambda }\} | a^{\dagger}_{\vec m} a_{{\vec m}'} |
      M' \{ n_{\vec \kappa} \}\rangle  \nonumber \\
&&\qquad \qquad - {p_M\over d(M,N_A)} \sum_{\vec \lambda} 
      \langle M \{ n_{\vec
      \lambda} \} | a^{\dagger}_{\vec n} a_{{\vec n}'}
      a^{\dagger}_{\vec m} a_{{\vec m}'} | M \{ n_{\vec \lambda} \}
      \rangle \Biggr ) \ .
\end{eqnarray}
We note that both terms on the right--hand side of Eq.~(\ref{eq20})
vanish unless we have ${\vec m} = {\vec n}'$ and ${\vec n} = {\vec
m}'$.  Moreover, the terms with ${\vec m} = {\vec n}$ cancel, so the
case ${\vec m} = {\vec n}$ is excluded from the summation. We also
note that in the sum over $M'$, all terms vanish but the one for which
we have $M' = M + \alpha$, with $\alpha$ defined in
Eq.~(\ref{eq6}). The sum $\sum_{\vec \lambda} | M \{ n_{\vec \lambda}
\} \rangle \langle M \{ n_{\vec \lambda} \} |$ is equal to the
projector onto the subspace of states with unperturbed energy $ M \
\hbar\nu$. The operator $a^{\dagger}_{\vec m} a_{{\vec n}}$ acting
upon states with energy $M' \hbar\nu$ can anyway only populate states
in this subspace. Therefore, the projector can be replaced by the unit
operator. As a result, the term multiplying $p_{M'}$ in
Eq.~(\ref{eq20}) becomes equal to $\sum_{{\vec \kappa}} \langle M' \{
n_{\vec \kappa}\} | a^{\dagger}_{\vec n} a_{{\vec m}}
a^{\dagger}_{\vec m} a_{{\vec n}} | M' \{ n_{\vec \kappa} \}
\rangle$. Divided by $d(M', N_A)$, this term is nothing but the mean
value of $n_{\vec n} [n_{\vec m} + 1]$ taken over the states with
fixed $M'$. We denote this quantity by $\langle n_{\vec n} [n_{\vec m}
+ 1] \rangle_{M+\alpha}$. The same argument is applied to the term in
Eq.~(\ref{eq20}) which multiplies $p_M$ and yields $\langle n_{\vec n}
[n_{\vec m} + 1] \rangle_M$. As a result, the master equation takes
the form of a rate equation,
\begin{equation}
\label{eq21}
\frac{{\rm d} p_M}{{\rm d}t} = 2 \sum_{{\vec m} \neq {\vec n}}
      \Gamma^{{\vec m},{\vec n}}_{{\vec n},{\vec m}} \biggl ( p_{M +
      \alpha} \langle n_{\vec n} [n_{\vec m} + 1] \rangle_{M+\alpha}
      - p_M \langle n_{\vec n} [n_{\vec m} + 1] \rangle_M \biggr ) \ . 
\end{equation}
It is straightforward to check that $\sum_M {\rm d}p_M(t)/{\rm d}t = 0$.
In Eq.~(\ref{eq21}), the rate coefficients $\Gamma$ and the
expectation values of the number operators are input quantities
defined in the framework of our model, and the $p_M$'s are the unknowns.

For later applications of Eq.~(\ref{eq21}), it is useful to realize
that the expectation values depend only upon the single--particle
energies and not on the single--particle wave functions. Therefore, we
define
\begin{equation}
\label{eq21a}
\bar{\Gamma}_{j,i} = \sum_{\vec{m}\ne
\vec{n}}\Gamma_{\vec{n},\vec{m}}^{\vec{m},\vec{n}}
\,\delta_{j\hbar\nu, \varepsilon(\vec{m})}
\,\delta_{i\hbar\nu, \varepsilon(\vec{n})}
\ .
\end{equation}
Then, Eq.~(\ref{eq21}) takes the form
\begin{equation}
\label{eq21b}
{{\rm d} p_M \over {\rm d}t} = 2 \sum_{\alpha=-K}^K
\sum_{j=\max{(0,-\alpha)}}^{\min{(K,K-\alpha)}} 
\bar{\Gamma}_{j,j+\alpha}\left[ p_{M+\alpha}
\langle n_{j+\alpha} (n_j+1) \rangle_{M+\alpha}
     - p_M \langle n_{j+\alpha} (n_j+1) \rangle_M \right]
\ .
\end{equation}
It is the advantage of this equation that the rates $\bar{\Gamma}$
can be calculated once and tabulated; the solution of
Eq.~(\ref{eq21b}) then requires a smaller number of steps than that of
Eq.~(\ref{eq21}).

To use Eq.~(\ref{eq21}), it is necessary to assign values to the
microcanonical averages $\langle n_{\vec n} [n_{\vec m} + 1]
\rangle_{M}$. We
do so by replacing the microcanonical average by the canonical
average, determining the temperature $T(E)$ by the thermodynamic
relation 
\begin{equation}
\label{eq22}
{1\over T(E)} = \frac{{\rm d}}{{\rm d}E}\,\kappa\ln{d(E,N_A)}
\end{equation} 
where $d(E,N_A)$ is the number of microstates of the Boson system $A$
at energy $E$, a smoothed version of the degeneracy $d(M, N_A)$. The 
microcanonical averages $\langle n_{\vec n} [n_{\vec m} + 1]
\rangle_{M}$ and $\langle n_{\vec m} [n_{\vec n} + 1] \rangle_M$ are
then approximated by $n_{\vec n}(T(E)) [n_{\vec m}(T(E)) + 1]$ and
$n_{\vec m}(T(E)) [n_{\vec n}(T(E)) + 1]$, respectively, with
\begin{equation}
\label{eq23}
n_{\vec m}(T(E)) = \frac{z(E) \exp[-\beta(E) \varepsilon({\vec m})]}{1 -
  z(E) \exp[-\beta(E) \varepsilon({\vec m})]} \ ,
\end{equation}
$\beta(E) = 1 / (\kappa T(E))$, and $z(E)$ determined from the total
number $N_A$ of particles in system $A$. The approximation neglects
possible correlations between the occupation numbers $n_{\vec n}$ and
$n_{\vec m}$ in the states labelled $M$. It is, thus, related to a
similar approximation used in Section~\ref{fac} where such
correlations will be discussed.

To carry out the calculation, we need to determine the
(energy--smoothed) number of accessible microstates $d(E,N_A)$ of
system $A$. We do so for $H'_{A-A} = 0$. We observe that $M$ may
attain very large values, and that the degeneracy $d(M, N_A)$ increases
rapidly (nearly exponentially) with increasing $M$. If, for instance,
the trap has 30 bound states and the system $A$ consists of $10^4$
atoms, then $M$ assumes all integers between zero and $3 \cdot 10^5$.
The very large degeneracy attained for large values of $M$ causes the
level density to increase very strongly with energy. This increase is
expected to level off at a point where the majority of Bosons occupies
the highest available single--particle level.

\begin{figure}[h]
 \phantom{.}\vspace{0.1cm}
  \begin{center}
    \leavevmode
    \parbox{0.6\textwidth}
           {\psfig{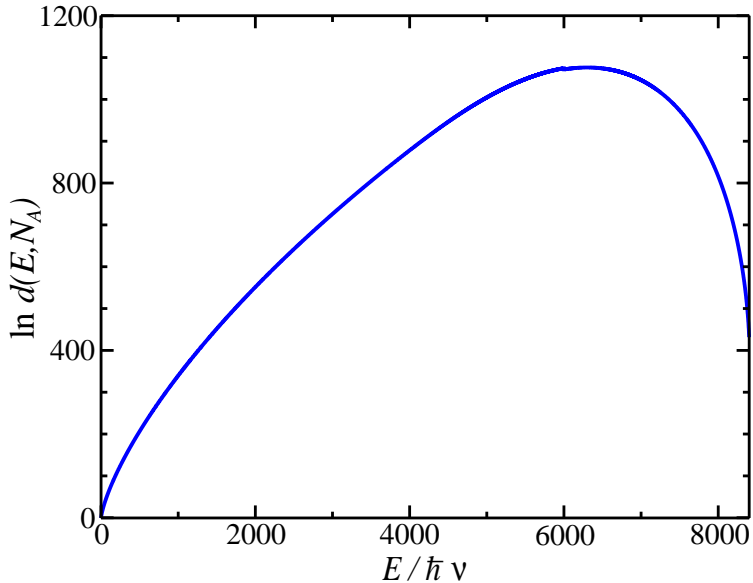}}
  \end{center}
\protect\caption{Entropy $\ln{d(E, N_A)}$ as a function of total
energy E for a system of $N_A = 400$ Bosons in a trap with highest 
single--particle orbital $K=21$.}
\label{fig1}
\end{figure}

We follow Ref.\cite{Holthaus} to compute $d(E,N_A)$. The number of
accessible microstates is related to the grand canonical partition
function $\Xi$ via
\ba
\Xi &\equiv& \prod_{j=0}^{K} \left(1 - z \, {\rm e}^{-\beta
  \varepsilon_j} \right)^{-g_j} \nonumber \\
&=& \sum_{N_A=0}^\infty z^{N_A} \sum_E{\rm e}^{-\beta E} \, d(E,N_A) \
,
\ea
where $K\hbar\nu$ is the energy of the highest
single--particle orbital in the harmonic trap, $E$ assumes values
$M\hbar\nu$, and $z$ denotes the fugacity. The degeneracy of
single--particle states with energies $\varepsilon_j=j\hbar\nu$ in the
harmonic trap is $g_j=(j+1)(j+2)/2$. We introduce $x\equiv {\rm
  e}^{-\beta\hbar\nu}$ and rewrite
\be
\label{grand}
\Xi = \sum_{N_A=0}^\infty z^{N_A} \sum_{M=0}^\infty x^M \, d(M, N_A) \
.
\ee
The microcanonical number of states  $d(E, N_A)$ can be obtained from
Eq.~(\ref{grand}) by contour integration, 
\be
\label{dE}
d(E,N_A) = {1 \over (2 \pi i)^2} \oint{\rm d}x\oint{\rm d} z \,
\exp{[-F(x,z)]} \ ,
\ee
with
\be
F(x,z) = \left({E \over \hbar \nu} + 1 \right) \ln x + (N_A + 1) \ln z
+ \sum_{j=0}^{K} g_j \, \ln{\left(1 - z \, x^j\right)} \ .
\ee
The integrations in Eq.~(\ref{dE}) can be performed using the
saddle--point approximation. However, care has to be taken in the
regime of Bose--Einstein--Condensation (BEC).  We recall that BEC is
reached in spherical harmonic traps at condensation temperatures
$\kappa T_c\approx 0.94 N_A^{1/3}\,\hbar\nu$, see
e.g. Ref.\cite{Stringari}. For $N_A = 400$ Bosons one thus finds
$\kappa T_c\approx 6.9\hbar\nu$. In the presence of BEC the
singularity due to the condensate must be excluded from the
saddle--point approximation \cite{Holthaus}. The saddle points of
Eq.~(\ref{dE}) are difficult to obtain for high energies, i.e.,
energies beyond the maximum of the curve in Fig.~\ref{fig1} and
especially for energies close to $N_A K\hbar\nu$. In this
regime we simply invert the single--particle spectrum and solve the
corresponding problem for low energies close to the (now highly
degenerate) ground state.

Fig.~\ref{fig1} shows the number of microstates $d(E, N_A)$ as a
function of the total energy $E$ for a system of $N_A = 400$ Bosons in
a trap with $K = 21$. The function $d(E,N_A)$ grows exponentially
for small energies and reaches a maximum; for energies beyond the
maximum the finite total number of single--particle orbitals causes a
decrease in $d(E, N_A)$. The number of accessible microstates is a
smooth function of energy. In the numerical computations we tabulate
$\ln d(E, N_A)$ for some energies and use these values for
interpolation.

In concluding this Section, we note that all steps in the derivation
leading to Eq.~(\ref{eq21}) apply equally if system $A$ consists of
Fermions instead of Bosons. This results in the following changes. We
replace $n_{\vec n}(T(E)) [n_{\vec m}(T(E)) + 1]$ everywhere by
$n_{\vec n}(T(E)) [1 - n_{\vec m}(T(E))]$. Moreover, Eq.~(\ref{eq23})
is replaced by $n_{\vec m}(T(E)) = z(E) \exp[-\beta(E)
\varepsilon({\vec m})]/[1 + z(E) \exp[-\beta(E) \varepsilon({\vec
  m})]]$. 

\section{Factorization}
\label{fac}

We turn to a different approximation to the master equation, written
in the form of Eqs.~(\ref{eq7},\ref{eq8}). As in Section~\ref{mic}, we
assume that $\rho_A(t)$ is diagonal in the $|M \mu \rangle$ basis, and
that -- for fixed $M$ -- Eq.~(\ref{eq19}) applies. We recall the basis
$| M \{ n_{\vec m} \} \rangle$ introduced in Section~\ref{mic}. We now
change the notation, omit the letter $M$, and impose no further
restrictions on the $n_{\vec m}$'s except for $\sum_{\vec m} n_{\vec
  m} = N_A$. In this basis, the reduced density matrix is diagonal,
and the master equation takes the form
\begin{eqnarray}
\label{eq24}
\frac{\rm d}{{\rm d}t} \langle \{ n_{\vec \kappa} \} | \rho_A(t) | \{
n_{\vec \kappa} \} \rangle &=& \sum_{{\vec m} \neq {\vec n}}
\Gamma^{{\vec m},{\vec n}}_{{\vec n},{\vec m}} \biggl ( 2 \langle \{
n_{\vec \kappa} \} | a^{\dagger}_{\vec m} a_{\vec n} \rho_A(t)
a^{\dagger}_{\vec n} a_{\vec m} | \{ n_{\vec \kappa} \} \rangle
\nonumber \\
&& - \langle \{ n_{\vec \kappa} \} | a^{\dagger}_{\vec n} a_{\vec m}
a^{\dagger}_{\vec m} a_{\vec n} \rho_A(t) | \{ n_{\vec \kappa} \}
\rangle \nonumber \\ 
&& - \langle \{ n_{\vec \kappa} \} | \rho_A(t) a^{\dagger}_{\vec n}
a_{\vec m} a^{\dagger}_{\vec m} a_{\vec n} | \{ n_{\vec \kappa} \}
\rangle \biggr ) \ .
\end{eqnarray}
The last two terms are equal and combine to $- 2 n_{\vec n}
[n_{\vec m} + 1] \langle \{ n_{\vec \kappa} \} | \rho_A(t) | \{
n_{\vec \kappa} \} \rangle$. Here, $n_{\vec n}$ and $n_{\vec m}$
denote the values of the occupation numbers for the single--particle
states ${\vec n}$ and ${\vec m}$, respectively, in the set $\{ n_{\vec
  \kappa} \}$. The first term yields $+ 2 n_{\vec m} [n_{\vec n} + 1]
\langle \{ \ldots (n_{\vec m} - 1) \ldots (n_{\vec n} + 1) \ldots \} |
\rho_A(t) | \{ \ldots (n_{\vec m} - 1) \ldots (n_{\vec n} + 1) \ldots
\} \rangle$. The notation indicates how the occupation numbers in the
set $\{ n_{\vec \kappa} \}$ are modified. The master equation becomes
\begin{eqnarray}
\label{eq25}
&& \frac{\rm d}{{\rm d}t} \langle \{ n_{\vec \kappa} \} | \rho_A(t) | \{
n_{\vec \kappa} \} \rangle = 2 \sum_{{\vec m} \neq {\vec n}}
\Gamma^{{\vec m},{\vec n}}_{{\vec n},{\vec m}} \nonumber \\
&& \qquad \times \biggl ( n_{\vec m} [n_{\vec n} + 1]  \langle
\{ \ldots (n_{\vec m} - 1) \ldots (n_{\vec n} + 1) \ldots \} |
\rho_A(t) | \{ \ldots (n_{\vec m} - 1) \ldots (n_{\vec n} + 1) \ldots
\} \rangle \nonumber \\
&& \qquad \qquad \qquad - n_{\vec n} [n_{\vec m} + 1] \langle \{
n_{\vec \kappa} \} | \rho_A(t) | \{ n_{\vec \kappa} \} \rangle \biggr
) \ .
\end{eqnarray}
We now take a partial trace of Eq.~(\ref{eq25}), summing over all
occupation numbers $n_{\vec m}$ of single--particle states ${\vec m}$
with ${\vec m} \neq {\vec m}_0$, keeping the latter fixed. The
corresponding partial trace of $\rho_A(t)$ is denoted by $p(n_{{\vec
    m}_0}, t)$. This yields
\begin{eqnarray}
\label{eq26}
\frac{\rm d}{{\rm d}t} p(n_{{\vec m}_0}, t) &=& 2 \sum_{{\vec m} \neq
  {\vec n}} \Gamma^{{\vec m},{\vec n}}_{{\vec n},{\vec m}}
  \Biggl ( \sum_{n_{\vec m} n_{\vec n}} n_{\vec m} [n_{\vec n} + 1]
  p(n_{{\vec m}_0}, n_{\vec m} - 1, n_{\vec n} + 1, t) \nonumber \\
&& \qquad - \sum_{n_{\vec m} n_{\vec n}} n_{\vec n} [n_{\vec m} + 1]
  p(n_{{\vec m}_0}, n_{\vec m}, n_{\vec n}, t) \Biggr ) \ .
\end{eqnarray}
The quantities $p(n_{{\vec m}_0}, n_{\vec m}, n_{\vec n}, t)$ are
defined as partial traces over $\rho_A(t)$ excluding the
single--particle states labelled ${{\vec m}_0}, {\vec m}$, and
${\vec n}$. It is easy to see that on the right--hand side of
Eq.~(\ref{eq26}), all terms cancel for which both ${\vec m}$ and
${\vec n}$ differ from ${{\vec m}_0}$. As a result, we find
\begin{eqnarray}
\label{eq27}
\frac{\rm d}{{\rm d}t} p(n_{{\vec m}_0}, t) &=& 2 \sum_{{\vec n} \neq
  {\vec m}_0} \Gamma^{{\vec m}_0,{\vec n}}_{{\vec n},{\vec m}_0}
  \biggl ( n_{{\vec m}_0} \sum_{n_{\vec n}} [n_{\vec n} + 1] \
  p(n_{{\vec m}_0} - 1, n_{\vec n} + 1, t) \nonumber \\
&& \qquad - [n_{{\vec m}_0} + 1] \sum_{n_{\vec n}}
  n_{\vec n} \ p(n_{{\vec m}_0}, n_{\vec n}, t) \biggr ) \nonumber \\
&& + 2 \sum_{{\vec m} \neq {\vec m_0}} \Gamma^{{\vec m},{\vec
  m}_0}_{{\vec m}_0,{\vec n}} \biggl ([ n_{{\vec m}_0} + 1]
  \sum_{n_{\vec m}} n_{\vec m} \ p(n_{{\vec m}_0} + 1, n_{\vec m} - 1,
  t) \nonumber \\
&& \qquad - n_{{\vec m}_0} \sum_{n_{\vec m}} [n_{\vec n} + 1]  \
  p(n_{{\vec m}_0}, n_{\vec n}, t) \biggr ) \ .
\end{eqnarray}
Except for the assumptions stated at the beginning of this Section,
all our steps have been exact. Unfortunately, the Eqs.~(\ref{eq27})
are not closed. We close them by introducing a factorization
assumption for the partial traces of the reduced density matrix. We
write
\begin{equation}
\label{eq28}
p(n_{\vec m}, n_{\vec n}, t) = p(n_{\vec m}, t) \ p(n_{\vec n}, t) \ .
\end{equation}
We discuss the implications of this assumption below and first deduce
the results. For clarity, we replace the symbols $p(n_{\vec m}, t)$ by
$p_{\vec m}(n, t)$. This indicates more clearly that $p_{\vec   m}(n,
t)$ is the density matrix for the single--particle state ${\vec m}$
with occupation number $n$. Eq.~(\ref{eq27}) takes the form
\begin{eqnarray}
\label{eq29}
\frac{\rm d}{{\rm d}t} p_{{\vec m}_0}(n, t) &=& 2 \sum_{{\vec n} \neq
  {{\vec m}_0}} \Gamma^{{{\vec m}_0},{\vec n}}_{{\vec n},{{\vec m}_0}}
  \Biggl ( n \sum_{n'} [n' + 1] \ p_{{\vec m}_0}(n - 1, t) \ p_{\vec
  n}(n' + 1, t) \nonumber \\
&& \qquad - [n + 1] \sum_{n'} n' \ p_{{\vec m}_0}(n, t) \ p_{\vec
  n}(n', t) \Biggr ) \nonumber \\
&& + 2 \sum_{{\vec m} \neq {{\vec m}_0}} \Gamma^{{\vec m},{{\vec
  m}_0}}_{{{\vec m}_0},{\vec m}} \Biggl ( [n + 1] \sum_{n'} n' \
  p_{{\vec m}_0}(n + 1, t) \ p_{\vec m}(n' - 1, t) \nonumber \\
&& \qquad - n \sum_{n'} [n' + 1] \ p_{{\vec m}_0}(n, t) \ p_{\vec
  m}(n', t) \Biggr ) \ .
\end{eqnarray}
Summing Eq.~(\ref{eq29}) over all $n$, we obtain ${\rm d} \sum_n
p_{{\vec m}_0}(n, t) / {\rm d}t = 0$. Because we have $\sum_n p_{{\vec
    m}_0}(n, t) = {\rm tr} \ \rho_A(t)$, this result is in keeping
with the condition ${\rm tr} \ \rho_A(t) = 1$. We define the average
occupation numbers $N_{\vec m}(t)$ in the single--particle state
${\vec m}$ at time $t$ by
\begin{equation}
\label{eq30}
N_{\vec m}(t) = \sum_n n \ p_{{\vec m}_0}(n, t) \ .
\end{equation}
Multiplying Eq.~(\ref{eq29}) by $n$ and summing over $n$, we obtain
the following closed system of rate equations for the $N_{\vec m}(t)$,
\begin{eqnarray}
\label{eq31}
\frac{\rm d}{{\rm d}t} N_{{\vec m}_0}(t) &=&  2 \sum_{{\vec n} \neq
  {{\vec m}_0}} \Gamma^{{{\vec m}_0},{\vec n}}_{{\vec n},{{\vec m}_0}}
  N_{\vec n}(t) \left( N_{{\vec m}_0}(t) + 1 \right) \nonumber \\
&& - 2 \sum_{{\vec m} \neq {{\vec m}_0}} \Gamma^{{\vec m},{{\vec
  m}_0}}_{{{\vec m}_0},{\vec m}} \left( N_{\vec m}(t) + 1 \right)
  N_{{\vec m}_0}(t) \ .
\end{eqnarray}
Eq.~(\ref{eq31}) constitutes the central result of this Section. It is
a set of rate equations for the mean occupation numbers $N_{\vec n}(t)$.
The form of these equations is intuitively obvious and might even have
been written down without much derivation. It is straightforward to
check that Eq.~(\ref{eq31}) implies the condition ${\rm d} \sum_n
N_{\vec n}(t) / {\rm d}t = 0$. Thus, the average particle number is
conserved.

The rate equations~(\ref{eq31}) have been derived under the tacit
assumption that a finite number of levels in the harmonic--oscillator
potential of the trap is available for occupation by atoms in system
$A$. In actual fact, the trap is open, and atoms with energies above a
critical energy are able to escape. This fact can most easily be
incorporated into Eq.~(\ref{eq31}) by restricting the values of ${\vec
  m}, {\vec n}$ and ${{\vec m}_0}$ to the bound states of the trap,
and by adding on the right--hand side of Eq.~(\ref{eq31}) a loss term.
This term allows for scattering into virtual harmonic--oscillator
levels which are actually unbound and from which the atoms can
escape. The derivation of the modified form of Eq.~(\ref{eq31}) is
straightforward and yields
\begin{eqnarray}
\label{eq31a}
\frac{\rm d}{{\rm d}t} N_{{\vec m}_0}(t) &=&  2 \sum_{{\vec n} \neq
  {{\vec m}_0}} \Gamma^{{{\vec m}_0},{\vec n}}_{{\vec n},{{\vec m}_0}}
  N_{\vec n}(t) \bigl ( N_{{\vec m}_0}(t) + 1 \bigr ) \nonumber \\
&& - 2 \sum_{{\vec m} \neq {{\vec m}_0}} \Gamma^{{\vec m},{{\vec
  m}_0}}_{{{\vec m}_0},{\vec m}} \bigl ( N_{\vec m}(t) + 1 \bigr )
  N_{{\vec m}_0}(t) \nonumber \\
&& - 2 \sum_{{\vec m}'} \Gamma^{{{\vec m}'},{{\vec m}_0}}_{{{\vec
  m}_0},{{\vec m}'}} N_{{\vec m}_0}(t) \ .
\end{eqnarray}
The summation over ${{\vec m}'}$ extends over the virtual states.

The rate coefficients obey the important thermodynamic identity
\begin{equation}
\label{eq32}
\Gamma^{{\vec m},{\vec n}}_{{\vec n},{\vec m}} = \Gamma^{{\vec
    n},{\vec m}}_{{\vec m},{\vec n}} \cdot \exp[ - \beta_B (
    \varepsilon_{\vec m} - \varepsilon_{\vec n} ) ] \ .
\end{equation}
This follows immediately from the definition in Eq.~(\ref{eq5}). For
the time--independent equilibrium solutions $N^{\rm eq}_{\vec m}$ of
Eq.~(\ref{eq31}), Eq.~(\ref{eq32}) implies the values
\begin{equation}
\label{eq33}
N^{\rm eq}_{\vec m} = \frac { z_A \exp[ - \beta_B \varepsilon_{\vec m}
  ] }{ 1 - z_A \exp[ - \beta_B \varepsilon_{\vec m} ] } \ ,
\end{equation}
with the fugacity $z_A$ determined from the condition $\sum_{\vec m}
N^{\rm eq}_{\vec m} = N_A$. 

It goes without saying that a parallel derivation applies if the atoms
in system $A$ are Fermions. The final rate equations are obtained by
replacing on the right--hand side of Eq.~(\ref{eq31}) the terms
$(N_{{\vec m}_0}(t) + 1)$ and $(N_{\vec m}(t) + 1)$ by $(1 - N_{{\vec
    m}_0}(t))$ and by $(1 - N_{\vec m}(t))$, respectively. The
equilibrium distribution is given by Eq.~(\ref{eq33}) with the minus
sign in the denominator replaced by a plus sign.

The central approximation Eq.~(\ref{eq28}) assumes that there are no
correlations between the occupancies of the states ${\vec m}$ and
${\vec n}$. The one correlation which must exist is total particle
number conservation. Thus, we expect that the approximation
Eq.~(\ref{eq28}) may fail whenever $N_{\vec m}$ or $N_{\vec n}$
approach $N_A$, the total number of particles in system $A$. For
Fermions, this can never happen. Thus, we focus attention on the case
of Bosons. A critical situation arises if the number $n_0$ of Bosons
in the ground state is comparable to $N_A$. A modification of our
previous derivation is, therefore, necessary only for those terms in
the master equation which contain the functions $p(n_0, n_{\vec n},
t)$. For these terms, we have previously assumed that $\sum_{n} n \
p(n_0, n_{\vec n}, t) = N_{\vec n}(t) \ p(n_0, t)$ with $N_{\vec n}(t)$
independent of $n_0$. We now improve on this approximation by letting
$N_{\vec n}(t)$ depend on $n_0$. This is accomplished by writing
$\sum_{n} n \ p(n_0, n_{\vec n}, t) = N_{\vec n}(n_0, t) \ p(n_0, t)$
and by using for $N_{\vec n}(n_0, t)$ the ansatz
\begin{equation}
\label{eq34} 
N_{\vec n}(n_0, t) = \frac{N_A - n_0}{N_A - N_{0}(t)} N_{\vec n}(t) \
,
\end{equation}
with $N_{0}(t) = \sum_{{n}_0} n_0 \ p(n_0, t)$ and $N_{\vec n}(t)$
independent of $n_0$. This approximation conserves particle number. It
leads to the following modification of the rate equations. We define
the variance $\delta N_0^2 = \sum_{n_0} (n_0)^2 p(n_0, t) - (
\sum_{n_0} n_0 p(n_0, t) )^2$ and have, omitting loss terms, for
${\vec m}_0 \neq 0$
\begin{eqnarray}
\label{eq35}
\frac{\rm d}{{\rm d}t} N_{{\vec m}_0}(t) &=&  2 \sum_{{\vec n} \neq
  {{\vec m}_0},0} \Gamma^{{{\vec m}_0},{\vec n}}_{{\vec n},{{\vec
  m}_0}} N_{\vec n}(t) \bigl ( N_{{\vec m}_0}(t) + 1 \bigr ) \nonumber
  \\
&& - 2 \sum_{{\vec m} \neq {{\vec m}_0},0} \Gamma^{{\vec m},{{\vec
  m}_0}}_{{{\vec m}_0},{\vec m}} \bigl ( N_{\vec m}(t) + 1 \bigr )
  N_{{\vec m}_0}(t) \nonumber \\
&& + 2 \Gamma^{{{\vec m}_0},0}_{0,{{\vec m}_0}} N_0(t) \left(
  N_{{\vec m}_0}(t) \left(1 - \frac{\delta N_0^2}{N_0(t)(N_A -
  N_0(t))}\right) + 1 \right) \nonumber \\
&& - 2 \Gamma^{0,{{\vec  m}_0}}_{{{\vec m}_0},0 } N_{{\vec m}_0}(t)
  \left( N_0(t) \left(1 - \frac{\delta N_0^2}{N_0(t)(N_A -
  N_0(t))}\right) + 1 \right) \ .
\end{eqnarray}
The equation for $N_0(t)$ has the form
\begin{eqnarray}
\label{eq36}
\frac{\rm d}{{\rm d}t} N_0(t) &=& + 2 \sum_{{\vec n} \neq 0}
  \Gamma^{0, {\vec n}}_{{\vec n}, 0} N_{\vec n}(t) \left(
  N_0(t) \left(1 - \frac{\delta N_0^2}{N_0(t)(N_A - N_0(t))}\right) +
  1 \right) \nonumber \\
&& - 2 \sum_{{\vec n} \neq 0} \Gamma^{{\vec n}, 0}_{0, {\vec n}}
  N_0(t) \left( N_{\vec n}(t) \left(1 - \frac{\delta N_0^2}{N_0(t)(N_A
  - N_0(t))}\right) + 1 \right) \ .
\end{eqnarray}
Comparing Eqs.~(\ref{eq35},\ref{eq36}) with the original rate
equations~(\ref{eq31}), we note the appearance of the correction term
$\delta N_0^2 / (N_0(t)(N_A - N_0(t)))$. This term effectively reduces
the coupling between the atoms in the ground state and the rest of the
gas and is expected to increase the cooling time.
Eqs.~(\ref{eq35},\ref{eq36}) are not closed as they stand and require
the solution of equations for higher moments of the occupation numbers.
These, in turn, would not be closed. We address this problem in
Eq.~(\ref{eq35a}) below. First, we present a simple estimate of the
correction term. We expect the term to be small both for $N_0(t) \ll
N_A$ and for $N_A - N_0(t) \ll N_A$. Thus, the correction term should
attain its maximum at or near $N_0(t) \approx (1/2) N_A$. It is then
reasonable to approximate the correction term by a simple smooth
function of $x = N_0 / N_A$ in the interval $[0,1]$, with very small
values at the end points and a maximum near the middle. Qualitative
support for this idea comes from Figures~1, 2, and 8 of
Ref.~\cite{koc00}. Using these Figures, we are led to the conclusion
that the correction term does indeed approximately have the form just
suggested, with a maximum value of the order of 1 per cent. It must be
stressed, of course, that Ref.~\cite{koc00} deals with equilibrium
phenomena and not, as we do here, with equilibration processes.

A quantitative evaluation of the correlation requires additional work.
To be brief, we only sketch the derivation of the coupled equations
which determine both, the mean occupation numbers $N_{\vec m}(t)$ and
the ground--state correlators. A simple derivation consists in
multiplying the master equation, Eqs.~(\ref{eq7},\ref{eq8}), either
with $a^{\dagger}_{\vec i} a_{\vec j}$ or with $a^{\dagger}_{\vec i}
a_{\vec j} a^{\dagger}_{\vec k} a_{\vec l}$ and taking the trace. We
use the assumptions introduced above. In particular, we assume that
${\rm tr} \ (a^{\dagger}_{\vec i} a_{\vec j} \rho_A)$ and ${\rm tr} \
(a^{\dagger}_{\vec i} a_{\vec j} a^{\dagger}_{\vec k} a_{\vec l}
\rho_A)$ are diagonal, and that the same assumption applies to the
terms involving six creation and annihilation operators. Such terms
result from the right--hand side of Eq.~(\ref{eq8}). For ${\vec k}
\neq 0$, we define the correlator $\delta_{\vec k}$ by writing ${\rm
  tr} \ (a^{\dagger}_0 a_0 a^{\dagger}_{\vec k} a_{\vec k} \rho_A(t))
= N_0(t) N_{\vec k}(t) + \delta_{\vec k}(t)$. To obtain a closed set
of equations, we use for $0 \neq {\vec k} \neq {\vec m} \neq 0$ the
approximation that ${\rm tr} \ (a^{\dagger}_0 a_0 a^{\dagger}_{\vec k}
a_{\vec k} a^{\dagger}_{\vec m} a_{\vec m} \rho_A(t)) = N_0(t) N_{\vec
  k}(t) N_{\vec m}(t) + N_{\vec k}(t) \delta_{\vec m} + N_{\vec m}(t)
\delta_{\vec k}$. This approximation amounts to the neglect of all
correlations not involving the ground state. As a result, we find that
Eqs.~(\ref{eq35},\ref{eq36}) retain their form, the terms $- N_0(t)
N_{{\vec m}_0}(t) \delta N_0^2 / \left( N_0(t)(N_A - N_0(t)) \right)$
and $- N_0(t) N_{\vec n}(t) \delta N_0^2 / \left( N_0(t)(N_A - N_0(t))
\right)$ being replaced by $\delta_{{\vec m}_0}$ and $\delta_{\vec n}$,
respectively. While Eqs.~(\ref{eq35},\ref{eq36}) suggest that we have
to determine a single function $\delta N_0^2 / \left( N_0(t)(N_A -
  N_0(t)) \right)$, the replacement just indicated shows that we must
determine a set of functions $\delta_{\vec k}$. The determining
equations for $\delta_{\vec k}$ with ${\vec k} \neq 0$ read
\ba
\label{eq35a}
\frac{\rm d}{{\rm d}t} \delta_{\vec k}(t) &=& 2 \left( N_{\vec k}(t) +
1 \right) \left( \sum_{{\vec m} \neq {\vec k},0} \Gamma^{{\vec k}, {\vec
m}}_{{\vec m}, {\vec k}} \delta_{\vec m} \right) + 2 \delta_{\vec k}
\left( \sum_{{\vec m} \neq {\vec k},0} \left( \Gamma^{{\vec k}, {\vec
m}}_{{\vec m}, {\vec k}} + \Gamma^{0, {\vec m}}_{{\vec m}, 0} \right)
N_{\vec m}(t) \right) \nonumber \\
&& - 2 N_{\vec k}(t) \left( \sum_{{\vec m} \neq {\vec k},0}
\Gamma^{{\vec m}, {\vec k}}_{{\vec k}, {\vec m}} \delta_{\vec m} \right)
- 2 \delta_{\vec k} \left( \sum_{{\vec m} \neq {\vec k},0} \left(
\Gamma^{{\vec m}, {\vec k}}_{{\vec k}, {\vec m}} + \Gamma^{{\vec m},
  0}_{0, {\vec m}} \right) [ N_{\vec m}(t) + 1 ] \right) \nonumber \\
&& + 2 \left( N_0(t) - N_{\vec k}(t) \right) \left( \Gamma^{0, {\vec
    k}}_{{\vec k}, 0} - \Gamma^{{\vec k}, 0}_{0, {\vec k}} \right)
[ \delta_{\vec k} + N_0(t) N_{\vec k}(t) ] \ .
\ea
The last term in Eq.~(\ref{eq35a}) is the ``feeding term'': For
$N_0(t) = 0$, the homogeneous equation has the solution $\delta_{\vec
  k}(t) = 0$. Deviations are due to non--zero occupation numbers of
the ground state.

\section{Consistency}
\label{con}

In Refs.~\cite{scu99,koc00}, the probability distribution for the
ground--state occupation for a system of Bosons coupled to a heat
bath was derived for the first time. In this Section, we show that 
our rate equations yield the same solution, although the system under
consideration differs from that of Refs.~\cite{scu99,koc00}. This
result then serves as consistency check for our derivation. Our
assumptions are similar to those used in Refs.~\cite{scu99,koc00}.
More specifically, we assume that the excited levels of system $A$ are
in thermal equilibrium with the heat bath (system $B$). This
assumption is quantified below.

We specialize Eq.~(\ref{eq27}) to ${\vec m}_0 = 0$,
\begin{eqnarray}
\label{eq37}
\frac{\rm d}{{\rm d}t} p_{0}(n, t) &=& 2 \sum_{{\vec n} \neq 0 }
  \Gamma^{{0},{\vec n}}_{{\vec n},{0}} \Biggl ( n \
  p_{0}(n - 1, t) \langle N_{\vec n}(t) \rangle_{n - 1} - [n + 1]  \
  p_{0}(n, t) \langle N_{\vec n}(t) \rangle_n \Biggr ) \nonumber
  \\ 
&& + 2 \sum_{{\vec m} \neq 0} \Gamma^{{\vec m},{0}}_{{0},{\vec m}}
  \Biggl ( [n + 1] \ p_{0}(n + 1, t) [\langle N_{\vec n}(t)
  \rangle_{n + 1} + 1] \nonumber \\
&& \qquad \qquad - n \ p_{0}(n, t) [ \langle N_{\vec n}(t)
  \rangle_n + 1] \Biggr ) \ .
\end{eqnarray}
Here, $\langle N_{\vec n}(t) \rangle_n = \sum_{n_{\vec n}} n_{\vec n}
\ p(n_0, n_{\vec n}, t) / p_0(n, t)$ is the expected value of $N_{\vec
  n}(t)$, given that there are $n$ Bosons in the ground state. We
define the cooling and heating coefficients
\ba
\label{eq38}
K_n = 2 \sum_{{\vec n} \neq 0 } \Gamma^{{0},{\vec n}}_{{\vec n},{0}}
\langle N_{\vec n}(t) \rangle_n, \ \
H_n = 2 \sum_{{\vec m} \neq 0} \Gamma^{{\vec m},{0}}_{{0},{\vec m}}
  [\langle N_{\vec m}(t) \rangle_n + 1] \ .
\ea
This yields
\begin{eqnarray}
\label{eq39}
\frac{{\rm d} p_0(n)}{{\rm d}t} &=& - \biggl \{ K_n \ (n + 1) p_{0}(n)
- K_{n - 1} \ n p_0(n - 1) \nonumber \\
&& \qquad + H_n \ n p_0(n) - H_{n + 1} \ (n + 1) p_0(n + 1) \biggr \}
\ .
\end{eqnarray}
Eqs.~(\ref{eq38},\ref{eq39}) agree formally with Eq.~(8) and the
definitions following it of Ref.~\cite{scu99}, and with
Eqs.~(20,21,22) of Ref.~\cite{koc00}.

The equilibrium solution of Eq.~(\ref{eq39}) has the form
\be
\label{eq40}
p_0(n) = p_0(0) \prod_{i=1}^n K_{i-1}/H_i \ .
\ee
Using this result for all $n = 0, \ldots, N_A$ and the constraint
$\sum_{n = 0}^N n p_0(n) = n_0$, we find for the normalization factor 
\be
\label{eq41}
p_0(0)^{-1} = Z = \frac{1}{n_0} \sum_{i = 0}^{N_A} \ i \prod_{j=1}^i
K_{j-1}/H_j
\ee
where $Z$ is the partition function.

We now assume that for ${\vec n} \neq 0$, the equilibrium mean
occupation numbers $\langle N_{\vec n} \rangle_n$ are given by a
thermal distribution,
\be
\label{eq42}
\langle N_{\vec n} \rangle_n = \frac{z_n \exp[ -\beta_B
  \varepsilon_{\vec n}]}{1 - z_n \exp[ -\beta_B \varepsilon_{\vec n}]}
  \ .
\ee
The fugacities $z_n$ are determined by the condition
\be
\label{eq43}
\sum_{{\vec n} \neq 0} \langle N_{\vec n} \rangle_n = N_A - n \ .
\ee

We use Eq.~(\ref{eq42}) and the thermodynamic identity
Eq.~(\ref{eq32}) and find that $H_n$ can be written in the form
\be
\label{eq44}
H_n = \exp[ \beta_B \varepsilon_0] \ 2 \sum_{{\vec n} \neq 0 }
  \Gamma^{{0},{\vec n}}_{{\vec n},{0}} \langle N_{\vec n}(t) \rangle_n
  / z_n \ .
\ee 
We assume that the temperature $T_B$ of the heat bath is so low that
$z_n \exp[ -\beta_B \varepsilon_{\vec n}] \ll 1$. Then, $H_n =H$
becomes independent of $n$, and the normalization condition
Eq.~(\ref{eq43}) implies that $z_n$ is approximately given by $z_n =
  (N_A - n) z$. The constant $z$ depends upon temperature but is
independent of $n$. Together with Eq.~(\ref{eq44}) and the
definition~(\ref{eq38}), this yields $K_n = (N_A - n) H z \exp[
  \beta_B \varepsilon_0 ]$. Inserting this into Eq.~(\ref{eq40}) and
using the normalization condition~(\ref{eq41}), we find
\be
\label{eq45}
p_0(n) = \frac{1}{Z} \frac{ \left( z \exp[\beta_B \varepsilon_0]
\right)^n }{(N_A - n)!} \ ,  
\ee
with
\be
\label{eq46}
Z =\frac{1}{n_0} \sum_{n = 0}^{N_A} n \frac{ \left( z \exp[\beta_B
  \varepsilon_0] \right)^n }{(N_A - n)!} \ .
\ee
The dependence of this result on $n$ coincides with that of the last
displayed equation on the left--hand side of page 023609 of
Ref.~\cite{koc00}. This shows the consistency of our result.

\section{Rate Coefficients}
\label{rat}

We turn to the computation of the rate coefficients defined in
Eq.~(\ref{eq5}); various quantities used in this equation are in turn
defined in Eqs.~(\ref{eq2}) to (\ref{eq4}) and Eq.~(\ref{eq6}). In
what follows we assume that the bath particles can accurately be
described by a Boltzmann distribution, see Eq.~(\ref{eq2}). This
assumption simplifies the computation of the rate coefficients
considerably since it leads to a factorization of the integrals
involved. Furthermore, it allows us to approximate in Eq.~(\ref{eq5})
the factor $[n(k')+1]$ by unity since the occupation numbers of the
bath states are small. We note that the $\tau$-integration in
Eq.~(\ref{eq5}) yields a $\delta$-function and thus implies
$\varepsilon(k) - \varepsilon(k') + \alpha \hbar \nu = 0$. We may
therefore replace in Eq.~(\ref{eq2}) $\varepsilon(k)$ by
$(\varepsilon(k) + \varepsilon(k') - \alpha \hbar \nu)/2$. We obtain
an integrand that is more symmetric in $k$ and $k'$. Our starting
point is thus
\ba
\Gamma^{{\vec m}, {\vec n}}_{{\vec n}, {\vec m}} &=& {n_B \Lambda_B
  \over 2 \hbar^2} \, {\rm e}^{+{1 \over 2} \beta_B \alpha \hbar \nu}
\int\limits_{-\infty}^\infty {\rm d} \tau \int {\rm d}^3k \, {\rm
  d}^3k' \, \gamma_{\vec{n}, \vec{m}} (\vec{k}, \vec{k'}) \,
\gamma_{\vec{m}, \vec{n}} (\vec{k'}, \vec{k}) \nonumber \\
&& \times \exp{ \left[ -\beta_B (\varepsilon(k) + \varepsilon(k')) /
  2\right]} \, \exp \left[ i \left (\varepsilon(k) - \varepsilon(k')
+ \alpha \hbar \nu \right) \tau / \hbar \right] \ .
\ea
We introduce the oscillator length $l_0 = \left({\hbar \over m\nu}
\right)^{1/2}$ and the dimensionless integration variables
$\vec{\kappa} = l_0\vec{k}, \vec{\kappa'} = l_0\vec{k'}, \vec{r} =
\vec{x}/l_0, \vec{r'} = \vec{x'}/l_0$, and $t = \tau \nu {m\over M}$.
We also define the dimensionless parameters $\alpha' = \alpha {M \over
m}$ and $\delta = {m \over M} \beta_B \hbar \nu$. Furthermore, we
introduce dimensionless oscillator wave functions without changing our
notation.  Choosing Cartesian coordinates for positions $\vec{r},
\vec{r}'$ and momenta $\vec{\kappa}, \vec{\kappa}'$ leads to a
factorization of the integrals,
\be
\label{Gamdimless}
\Gamma^{{\vec m}, {\vec n}}_{{\vec n}, {\vec m}} = 
\omega \, {\rm e}^{{1 \over 2} \alpha' \delta}
\int\limits_{-\infty}^{\infty} {\rm d}t \,
\exp{(i \alpha' t)} \prod_{j=x,y,z} I_{m_j,n_j}(t) \ .
\ee
The $t$-dependent integrals $I_{m,n}$ are given by
\ba
\label{Imn}
I_{m,n}(t) &=& \int\limits_{-\infty}^\infty {\rm d}r \, {\rm d}r' \,
{\rm d}\kappa \, {\rm d}\kappa' \, \psi_m(r) \, \psi_n(r) \,
\psi_m(r') \, \psi_n(r') \nonumber \\
&& \exp{ \left[ -{\delta \over 4}(\kappa^2 + \kappa'^2) \right]} \,
\exp{\left[i{t \over 2}(\kappa^2 - \kappa'^2) \right]} \, \exp{\left[
  -i(\kappa - \kappa')(r-r') \right]} \ .
\ea
The normalization factor $\omega$ in Eq.~(\ref{Gamdimless}) is given
by
\be
\label{omega}
\omega={1\over 32\pi^4}\Lambda_B^3 n_B 
{a^2\over l_0^2}{(M+m)^2\over M m}\,\nu \ .
\ee
Below we find that $I_{m,n}(t)$ is an even function in $t$ and
symmetric under exchange $m\leftrightarrow n$. Thus, the rate
coefficients (\ref{Gamdimless}) obviously fulfill the thermodynamic
identity (\ref{eq32}). Most of the parameter dependence for the rate
coefficients is simply contained in the normalization $\omega$, which
obeys $\omega \ll \nu$ in realistic applications. The nontrivial and
interesting parameters are $\delta$ and $\alpha'$. Below we discuss 
how the rate coefficients depend on these parameters.

The four integrations over $r, r', \kappa, \kappa'$ can be done exactly.
We first perform the integrations over $r, r'$ and then the resulting
Gaussian integrations over $\kappa, \kappa'$~\cite{lew95}. We find
\be 
\label{Imnex}
I_{n,m} = 2 \sqrt{\pi \over \delta} \, \sum_{k,l=0}^{\min{(m,n)}}
c_{m,n,l} \, c_{m,n,k} \,{\Gamma(m + n - k - l + 1/2) \over \left ( 1
  + {\delta/4} + {t^2 / \delta} \right)^{m + n - k - l + 1/2}} \ ,
\ee 
with coefficients 
\be
c_{m,n,l} = {(-1)^l \, \sqrt{m! \, n!} \over l! \, (m-l)! \, (n-l)!} \
.
\ee
Insertion of these results into Eq.~(\ref{Gamdimless}) leads to an
integral over $t$ of the form
\be
\label{Vq}
V_n(\alpha') \equiv \int\limits_{-\infty}^\infty {\rm d}t \,
{{\em e}^{i \alpha' t} \over \left ( 1 + {\delta \over 4} + {t^2 \over
    \delta} \right )^{n + 1/2}} \ .
\ee
This integral can be done exactly \cite{Gradshteyn}. We find
\be
V_n(\alpha') = 2{\sqrt{\delta \over \pi}} \, \Gamma(1/2-n) \,
\left ({-|\alpha'| \delta \over2 \sqrt{ \delta (1 + \delta/4)}}
\right)^n K_n \left( |\alpha'|\sqrt{ \delta(1 + \delta/4)}\right) \ .  
\ee
Here $K_n(x)$ denotes the modified Bessel function and $n$ is a
positive integer. We also need the value of $V_n(\alpha')$ for
$\alpha' = 0$. It is safe to take the limit $\alpha'\to 0$ in the
equation above. We find
\be
V_n(0)= \sqrt{\pi \delta} \, {\Gamma(n) \over \Gamma(n + 1/2)} \,
\left (1 + \delta/4 \right )^{-n} \ .
\ee
The exact integration of $V_n$ constitutes an improvement over the
approximate result given in Ref.~\cite{lew95}. We have reduced the
computation of the rate coefficients to a six-fold sum ($q_j=m_j + n_j
- l_j - k_j$ and $p_j = {\min{(m_j,n_j)}}$)
\be
\label{Gamfin}
\Gamma^{{\vec m}, {\vec n}}_{{\vec n}, {\vec m}} = 
8\left({\pi\over\delta}\right)^{3\over 2}\omega \,{\rm e}^{{1\over 2}\alpha'\delta}
\,\sum_{l_x,k_x=0}^{p_x}\, \sum_{l_y,k_y=0}^{p_y} \,
\sum_{l_z,k_z=0}^{p_z} \left ( \prod_{j=x,y,z} c_{m_j,n_j,l_j} \,
c_{m_j,n_j,k_j} \, \Gamma(q_j + 1/2) \right) V_{1 + q_x + q_y + q_z} \
. 
\ee
However, the numerical computation of the rate coefficients still
presents a formidable problem. This is due to the cancellations that
occur in the sum in Eq.~(\ref{Gamfin}). These cancellations arise
already in the two--fold sum in Eq.~(\ref{Imnex}). For $\min({m,n})$
exceeding $\approx 20$, the individual terms of the sum vary by so
many orders of magnitude that a numerical computation with double
precision floating point variables leads to a total loss of precision
in the final numerical result. Resorting to extended numerical
precision is no practical solution since it increases the computation
time by orders of magnitude. Such an increase is unaffordable in view
of the fact that realistically, traps contain bound states of to the
30th harmonic--oscillator level or so. The number of rate coefficients
needed in such a case is about $[(1/6) \cdot 30^3]^2/2\approx 10^7$. 

To circumvent this problem we go back to Eq.~(\ref{Imn}) and perform
the Gaussian integrals over $\kappa$ and $\kappa'$ first. This yields
\be
\label{int1}
I_{m,n} = 2 \pi \sqrt{2 \gamma \over \delta}
\int\limits_{-\infty}^\infty {\rm d}r \, \psi_n(r) \left (
\int\limits_{-\infty}^\infty {\rm d}r' \psi_m(r') \exp{ \left [
  -\gamma(r-r')^2 \right]} \, \psi_n(r') \right) \psi_m(r) \ .
\ee
We have used the shorthand notation 
\be
\label{gam}
\gamma = 2 \delta / (\delta^2 + 4 t^2) \ .
\ee 
The integral in Eq.~(\ref{int1}) can be viewed as the matrix element
of the Gaussian two--body ``interaction potential'' $G=\exp{ \left [
  -\gamma(r-r')^2 \right]}$ taken between a pair of two--body states,
i.e., 
\be 
\label{Imnsc}
I_{m,n} = 2 \pi \, \sqrt{2 \gamma \over \delta} \,\, \langle
\vec{\mu} | G | \vec{\nu} \rangle \ .
\ee 
We have denoted the states by their quantum numbers as $\vec{\mu} =
(n,m)$ and $\vec{\nu} = (m,n)$. We recall that we are particularly
interested in the values of these matrix elements for large values of
the integers $m,n \gg 1$. In the limit of large quantum numbers a
semiclassical evaluation of the matrix element is promising. Following
Ref.~\cite{Morehead} we use the semiclassical approximation and obtain
\be
\langle \vec{\mu} | G | \vec{\nu} \rangle \approx {1 \over ( 2 \pi)^2}
\int\limits_0^{2\pi}{\rm d} \vartheta \int\limits_0^{2\pi} {\rm d}
\vartheta' \, {\rm e}^{-i(\vec{\mu} - \vec{\nu}) \cdot
  \vec{\vartheta}} \, G \left( \vec{\vartheta}, {\vec{J}_{\vec{\mu}}
  + \vec{J}_{\vec{\nu}} \over 2} \right).
\ee 
Here, $\vec{J}_{\vec{\nu}}=\vec{\nu}$ and $\vec{J}_{\vec{\mu}} =
\vec{\mu}$ denote the vectors $(J,J')$ of classical actions $J$ of the
corresponding initial and final quantum states, respectively, and
$\vec{\vartheta} = (\vartheta, \vartheta')$. We use the
harmonic--oscillator relations $r = \sqrt{2J} \cos{\vartheta}$, $r' =
\sqrt{2J'} \cos{\vartheta'}$ between positions and angle--action
variables and obtain
\be
\langle \vec{\mu} | G | \vec{\nu} \rangle = {1 \over(2 \pi)^2}
\int\limits_0^{2\pi} {\rm d} \vartheta \int\limits_0^{2\pi} {\rm d}
\vartheta' {\rm e}^{-i(n-m)(\vartheta - \vartheta')} \, {\rm e}^{-
  \gamma(n + m)(\cos{\vartheta} - \cos{\vartheta'})^2} \ .
\ee
Transformation to coordinates $\phi = \vartheta - \vartheta',
\phi' = (\vartheta + \vartheta')/2$ allows us to perform one
integration analytically~\cite{Gradshteyn}. We arrive at
\be
\label{semicl}
\langle \vec{\mu} | G | \vec{\nu} \rangle = {1 \over \pi}
\int\limits_0^\pi {\rm d} \phi \,{ \rm e}^{-2 \gamma(n + m) \sin^2 {
    \phi\over 2} } \, I_0 \left(2 \gamma(n + m) \sin^2{\phi \over 2}
\right) \, \cos{[(n - m) \phi]} \ .
\ee
Here $I_0$ denotes the modified Bessel function of zeroth order. For
the computation of the rate coefficients we perform the remaining
integration over $\phi$ and, subsequently, the integration over $t$,
numerically.
 
Using Mathematica we checked that for $m,n \gg 1$, Eqs.~(\ref{Imnsc})
and (\ref{semicl}) are in excellent agreement with Eq.~(\ref{Imnex}).
This is exactly the regime we are interested in. As expected, the
semiclassical approximation becomes inaccurate for $|m-n| \approx
m+n$. This is of no concern to us, however, since our exact
expressions (\ref{Imnex}) and (\ref{Gamfin}) are easily computed
numerically in this regime. We conclude that our results permit us to
compute the rate coefficients efficiently and accurately for rather
large systems. 

\begin{figure}[h]
 $\phantom{.}$
  \begin{center}
    \leavevmode
    \parbox{0.75\textwidth}
           {\psfig{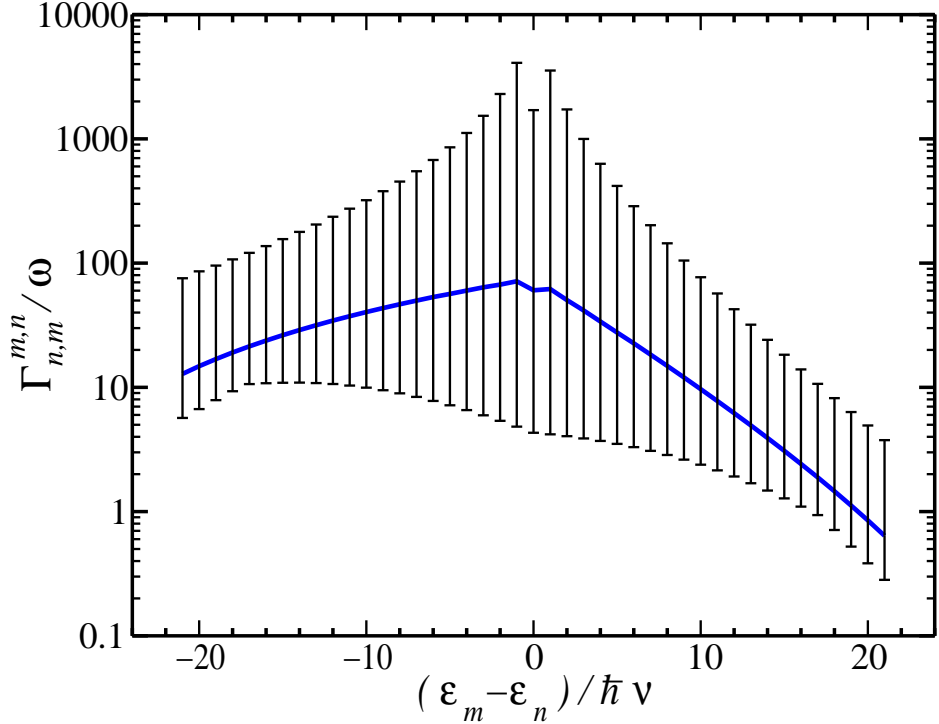}}
  \end{center}
\protect\caption{Distribution of scaled rate coefficients
$\Gamma^{\vec{m},\vec{n}}_{\vec{n},\vec{m}} / \omega$ for ${}^{23}$Na
atoms in a ${}^{87}$Rb bath at temperature $1/\beta_B = 7\hbar\nu$ as
a function of energy transfer $(\varepsilon(\vec{m}) - \varepsilon(
\vec{n})) / \hbar \nu$: Average value (line), maximal and minimal
values (error bars).}
\label{fig2}
\end{figure}

Fig.~\ref{fig2} shows the rate coefficients for ${}^{23}$Na atoms in a
${}^{87}$Rb bath at temperature $1/\beta_B = 7\hbar\nu$ as a function
of the energy transfer $(\varepsilon(\vec{m}) - \varepsilon(\vec{n}))
/ \hbar \nu$. The plotted line is the average value and the error bars
indicate maximal and minimal rate coefficients at given energy transfer.
In this example we choose the highest single--particle orbit at
excitation energy $K\hbar\nu$ with $K=21$. The rate coefficients attain
maximum values close to zero energy transfer. This fact supports the
approximation introduced in Eq.~(\ref{eq19}). We note that the
distribution of the rate coefficients at fixed energy transfer
displays a large variance. The number of individual rate coefficients
increases dramatically with decreasing modulus of energy transfer. The
asymmetry between transitions with negative energy transfer (cooling)
and positive energy transfer (heating) is due to the thermodynamic
identity~(\ref{eq32}). Approximately 2\% of the rate coefficients had
to be computed using the semiclassical method. This fraction increases
with increasing trap cutoff $K$.

For applications of our results, the dependence of the cooling rate on
the system parameters is of central interest. The main dependence of
the rate coefficients on the system parameters is given by the common
factor $\omega$ defined in Eq.~(\ref{omega}). This quantity sets the
overall time scale. As expected, $\omega$ is linear in both the
density of bath atoms and the oscillator frequency and quadratic in
the scattering length. It is a symmetric function of the masses of the
atoms in systems $A$ and $B$, attaining maximum values when one of the
two masses is much larger than the other. The dependende of the rate
coefficients on the parameters $\delta = (m/M) \beta_B \hbar \nu $ and
$\alpha' = (M/m) \alpha$ is implicit. Eq.~(\ref{Gamdimless}) expresses
the rate coefficients as Fourier transforms with respect to $t$ of a
three--fold product of functions $I_{m,n}$. These functions, given in
Eq.~(\ref{Imnex}) and (for the semiclassical approximation) in
Eq.~(\ref{Imnsc}) and Eq.~(\ref{semicl}), depend on $t$ and on the
parameter $\delta$.  We note that $I_{m,n}$ is an even function of $t$;
we restrict the discussion to non--negative arguments. We found that
the $I_{m,n}$'s are positive and monotonically decreasing functions of
$t$ which vanish asymptotically for $t \to \infty$. For fixed $\delta$
and fixed sum $m + n$ the functions $I_{m, n}$ increase with decreasing
$|m - n|$. This behavior is unaffected by the integration over $t$ and
is clearly reflected in Fig.~\ref{fig2}. For fixed $\delta$ and fixed
energy transfer $|m - n|$ but increasing values of $m + n$, the values
of the functions $I_{m,n}$ increase at the origin but fall off more
quickly with increasing $t$. In realistic applications we have
${\beta_B \hbar \nu \ll 1}$, and this results in $\delta\ll 1$ unless
$m \gg M$. For fixed $m,n$, the function $I_{m,n}$ increases with
decreasing $\delta$. However, this increase does not translate
directly into an increase of the rate coefficients. For large energy
transfer $|m-n|$ the increase is over--compensated by the highly
oscillating exponential in the integrating over $t$, and the cooling
process evolves mainly over transitions to levels that are close in
energy. We also observe that an increase of the ration $M/m$ decreases
$\delta$ but increases $\alpha'$ and, thus, the frequency of the
oscillations of the exponential.

We computed the distribution of rate coefficients also for a system of
${}^{87}$Rb atoms in a bath of ${}^{87}$Rb atoms at temperature
$1/\beta_B = 7\hbar\nu$. (One may assume that system and bath atoms are
in different hyperfine states.) When comparing the distribution to
the case of ${}^{23}$Na atoms in a ${}^{87}$Rb bath depicted in
Fig.~\ref{fig2} we find the following: At small energy transfer the
Rb-Rb distribution has average and maximal values that are about three
times smaller than for Na-Rb. However, the averages and maximal values
of the Rb-Rb system decrease less fast with increasing modulus of the
energy transfer and are similar to those of Na-Rb at maximal energy
transfer.

\section{Numerical Results}
\label{num}

In this Section we study the cooling process for Bosons by
solving the rate equations~(\ref{eq21b}) and (\ref{eq31}) numerically.
We compare the results. As an example we take system $A$ to be
composed of $N_A = 400$ ${}^{23}$Na atoms, while the ${}^{87}$Rb bath
has the temperature $1/\beta_B = 7\hbar \nu$. This temperature is
slightly above the condensation temperature $\kappa T_c = 6.93 \hbar
\nu$ for harmonic traps. We assume that the highest bound state in the
harmonic trap has energy $K \hbar \nu$ where $K = 21$.

In the microcanonical approach, the rate equations are given by
Eq.~(\ref{eq21b}). The computation of the input parameters $\langle
n_j (n_{j - \alpha} + 1) \rangle_M$ and $\bar{\Gamma}_{j,i}$ is
described in Section~\ref{mic} and Section~\ref{rat}, respectively.
We checked that the computed expectation values $\langle n_j
\rangle_M$ fulfill the sum rules $\sum_j {1\over 2} (j+1) (j+2) \langle
n_j \rangle_M = N_A$ and $\sum_j {1\over 2} j (j+1) (j+2) \langle n_j
\rangle_M = M$ to 1\% accuracy. The homogeneous system of linear
equations given by Eq.~(\ref{eq21b}) can be put into matrix form. We
introduce the sparse matrix $A$ which has $2K + 1$ non--vanishing
elements in each row and column and write Eq.~(\ref{eq21b}) somewhat
symbolically as ${\rm d}p_M / {\rm d}t = \sum_N A_{M,N} p_N$. We note,
however, that in our example the matrix $A$ has dimension 8401. With
$\max{(M-K,0)} \le N \le \min{(M+K, K N_A)}$, the off--diagonal
elements of $A$ are given by
\be
A_{M,N} = 2 \sum_{j = \max{(0,M-N)}}^{\min({K,K+M-N})}
\bar{\Gamma}_{j,j+N-M} \, \langle n_{j + N - M} (n_j + 1) \rangle_N \ ,
\ee
and the diagonal elements are
\be
A_{M,M}=-2\sum_{\alpha = \max{(-K,M-K N_A)}}^{\min{(K,M)}}
\, \, \sum_{j = \max{(0,-\alpha)}}^{\min{(K,K - \alpha)}}
\bar{\Gamma}_{j,j + \alpha} \langle n_{j+\alpha} (n_j+1)\rangle_M \ .
\ee
The limits on the summations result either from Eq.~(\ref{eq21b}) or
are a consequence of the identities $\langle n_{i} (n_j + 1) \rangle_M
= 0$ for $i > M$ and $\langle n_i(n_j + 1) \rangle_{K(N_A - 1) + l} =
0$ for $i < l$. The conservation of probability in Eq.~(\ref{eq21b})
is manifest since the elements in each column of $A$ sum up to zero.
The complete diagonalization of the matrix $A$ is expensive and
unnecessary. Instead, we compute only the eigenvalues with the largest
real parts. For stability reasons, there are no eigenvalues with
positive real parts. The equilibrium solution is determined by the
zero eigenvalue. The equilibration rate is thus equal to the modulus
of the real part of the eigenvalue with largest negative real part; the
corresponding mode is damped out last. The eigenvalues of interest are
computed using the sparse matrix solver ARPACK\cite{Arpack}. Within
our numerical accuracy we found one zero eigenvalue and no eigenvalues 
with positive real parts. For the equilibration rate we found
$\gamma_{\rm eq}\approx 2.7\times 10^4\omega$, with $\omega$ defined
in Eq.~(\ref{omega}). The equilibrium energy $E_{\rm eq} = 3961.3$ can
easily be computed from the eigenvector belonging to eigenvalue zero;
alternatively, it can be obtained from the probability distribution
for large times in a numerical integration of the rate
equations~(\ref{eq21b}).

\begin{figure}[h]
 $\phantom{.}$
  \begin{center}
    \leavevmode
    \parbox{0.75\textwidth}
           {\psfig{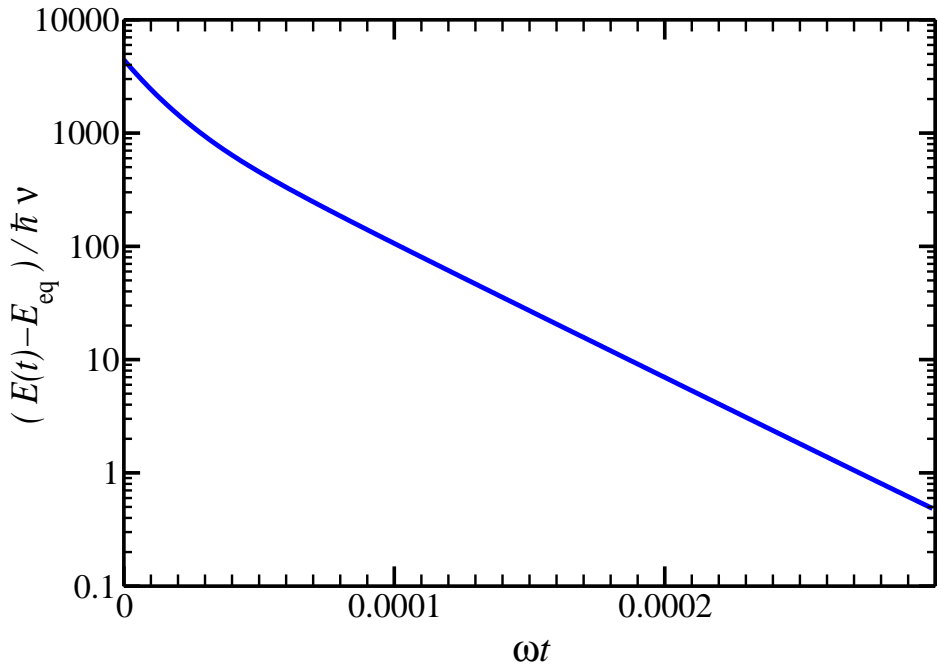}}
  \end{center}
\protect\caption{Energy difference $E(t)-E_{\rm eq}$ as a function of time 
within the microcanonical approach in the Na-Rb system.}
\label{fig3}
\end{figure}

It is also interesting to follow the time evolution of the total
energy $E(t)=\hbar \nu \sum_M M p_M(t)$ for a system of Na atoms in a
Rb bath. As the initial condition we take $p_{K N_A} = 1$ and vanishing
values for all other probabilities. Fig.~\ref{fig3} shows the time
evolution of $E(t)$ as obtained upon integration of the rate equations.
Initially, the decay is fast but non--exponential. This is due to the
fact that many eigenvalues of the matrix $A$ contribute to the cooling
process. At later times the equilibrium value $E_{\rm eq}$ is
approached exponentially fast with the rate $\gamma_{\rm eq}$ given
above. Inspection of the data used in Fig.~\ref{fig1} shows that the
temperature corresponding to the equilibrium energy $E_{\rm eq}$ is
$\kappa T=7.0 \hbar\nu$. This agrees well with the temperature of the
bath.

\begin{figure}[h]
 $\phantom{.}$
  \begin{center}
    \leavevmode
    \parbox{0.75\textwidth}
           {\psfig{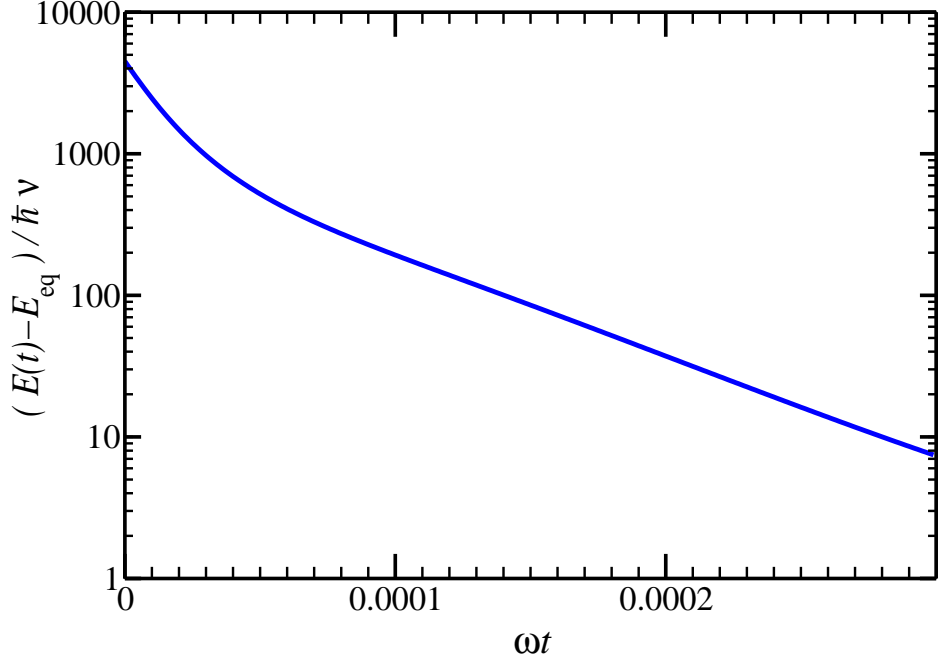}}
  \end{center}
\protect\caption{Energy difference $E(t)-E_{\rm eq}$ as a function of 
time in the approach based on factorization in the Na-Rb system.}
\label{fig4}
\end{figure}

We turn to the description of the cooling process in terms of the rate
equations~(\ref{eq31}). The number of coupled nonlinear differential
equations defined by Eq.~(\ref{eq31}) is given by the number of
single--particle states in the trap and equals 2024 in our example.
Again, we consider Na atoms in a Rb bath. As the initial condition we
take the $N_A = 400$ Bosons to be equally distributed over the
degenerate single--particle orbitals with energy $K\hbar \nu$ while
all other orbitals are empty. This initial condition corresponds to
the situation discussed above for the microcanonical approach.
Fig.~\ref{fig4} shows a plot of the total energy as a function of time.
Initially, the decay is fast but nonexponential and can barely be
distinguished from the decay within the microcanonical approach
(compare with Fig.~\ref{fig3}). At later times the equilibrium $E_{\rm
  eq}\approx 3901$ is approached exponentially fast with the
equilibration rate $\gamma_{\rm eq}=1.6\times 10^4\omega$. This rate
was determined from the time evolution shown in Fig.~\ref{fig4}.
Alternatively, the rate might be obtained upon linearization of the
rate equations~(\ref{eq31}) around the equilibrium solution of
Section~\ref{con}. However, we did not pursue this point any further.
The equilibrium energy is about 2\% smaller than the corresponding
value found in the microscopic approach. According to the data used in
Fig.~\ref{fig1}, we find a temperature of $\kappa T=6.9\hbar\nu$,
slightly deviating from the bath temperature. We recall that the bath
temperature enters the rate equations through the rate coefficients.
Within our numerical accuracy, and especially because of the
semiclassical approximation, we expect the relative error of the rate
coefficients to be a few percent. In view of this fact, the agreement
between the temperature of the bath and the one found in our
calculation, is quite satisfactory.

Comparing the results obtained in the microcanonical approach and by
using factorization, we note that the equilibration rate obtained from
the microcanonical approach is about a factor $\approx 1.7$ larger
than the corresponding rate found using factorization. The loss of
energy at short times is, however, practically identical in both
approaches. It is evident from Fig.~\ref{fig3} and Fig.~\ref{fig4}
that most of the energy is removed from system $A$ during the early
period of the cooling process. Therefore, both approaches yield
comparable predictions for the cooling time. For example, about 90\%
of the finally removed energy $E(t=0)-E_{\rm eq}$ have been transfered
to the bath at a time $t_{\rm cool}\approx 0.6\times
10^{-4}\omega^{-1}$.

We also note that the results obtained in the microcanonical
approach are somewhat less sensitive to the exact values of the
rate coefficients than those obtained using factorization. This is
plausible since only sums over many rate coefficients enter the rate
equations in the former case.
 
Let us finally discuss sympathetic cooling of systems with equal
masses.  Note that the regime of similar masses lies somewhat
outside the scope of this work. We recall that the bath particles are
assumed to be much heavier than the particles of the system. This
difference in masses is also reflected by the use of different wave
functions for the particles of the bath and the particles of the
system. A detailed computation with identical wave functions for
system and bath may thus yield quantitatively different results. With
these cautionary remarks in mind we turn to sympathetic cooling of
${}^{87}$Rb atoms in a bath of ${}^{87}$Rb atoms. (System and bath
atoms are assumed to be in different hyperfine states). The results
apply, of course, equally to all other cases where cooled atoms and
bath particles have equal masses. Within the microcanonical approach
we found equilibration to a final state with energy $E_{\rm eq}=3965.9
\hbar \nu$. This corresponds to the temperature $\kappa T = 7.0\hbar
\nu$ and agrees well with the temperature of the bath. The
equilibration rate is $\gamma_{\rm eq} = 1.1 \times 10^4 \omega$ and
the cooling time $t_{\rm cool} \approx 1.5 \times 10^{-4}
\omega^{-1}$.  Using the approach based on factorization we found for
the energy of the equilibrated system $E_{\rm eq} \approx
3900\hbar\nu$ corresponding to a temperature $\kappa T \approx 6.9
\hbar \nu$. The equilibration rate was $\gamma_{\rm eq} \approx 0.6
\times 10^4\omega$; the cooling time $t_{\rm cool} \approx 1.5 \times
10^{-4}\omega^{-1}$.  At the initial stages of the cooling process
there is again barely any difference between the microcanonical
approach and the one based on factorization. Nevertheless,
equilibration rates vary by a factor of about 1.7. Comparing the
results for the Rb-Rb system with the Na-Rb system one thus finds that
sympathetic cooling times (in units of $\omega^{-1}$) decrease
significantly with decreasing mass ratio $m/M$ between system and bath
particles. In our case, the factor (about 2.5) is similar to the
factor reported at the end of Section~\ref{rat}.

\section{Conclusions}
\label{conc}

We have derived rate equations for the sympathetic cooling of systems
composed of Bosons or of Fermions. The rate equations were obtained
from the master equation derived in Ref.~\cite{lew95} in a sequence of
steps. First we used the perturbatively weak interaction $H_{A-A}'$
between the particles in the system to be cooled to lift the
degeneracy of the many--body states in the trap. This allowed us in a
second step to use the decoherence argument and resulted in a reduced
density matrix of diagonal form. In a third step we invoked an
ergodicity argument and assumed that the equilibration between
(quasi--)degenerate states is much faster than the cooling process;
this assumption reduced the number of independent diagonal elements
further. From here on we proceeded along two different routes. Within
the microcanonical approach we obtained rate equations that govern the
occupation probabilities of sets of (quasi--)degenerate many--body
states. This results in a linear problem with sparse matrix. The
dimension of the problem increases with particle number and trap
cutoff. Alternatively, we traced the reduced density matrix over
single--particle states. Assuming that occupation probabilities for
different orbitals factorize we obtained a nonlinear set of rate
equations for the mean occupation probabilities of the
single--particle orbitals. The dimension of this problem depends only
on the trap cutoff. We showed how to extend this approach to include
correlations and particle escape from the open trap, although we did
not yet test these extensions numerically.

We provided several checks on our assumptions. To check the
consistency of our assumptions we used the rate equations and computed
the probability distribution for the ground state occupation; the
results agree with those of the literature. We further showed that the
decoherence argument is justified. Off--diagonal elements of the
reduced density matrix decay on a time scale that is inversely
proportional to the square root of the number of bath particles and,
thus, very short. The ergodicity argument was supported by the
dependence of the rate coefficients on energy transfer, see
Fig.~\ref{fig2}.

Any solution of the rate equations requires as input the rate
coefficients. We derived analytical expressions for these coefficients.
A numerical evaluation of the resulting formulas becomes impractical,
however, for transitions between high--lying single--particle states
that are close in energy. We solved this problem by using a
semiclassical approximation. We note that the computation of the rate 
coefficients is the time--consuming part in practical applications of
the rate equations derived in this work. We discussed the dependence
of the rate coefficients on the parameters of the system.

The two different rate equations obtained in this paper yield results
that are in semi--quantitative agreement. Within both approaches, the
cooling times are about the same: Both descriptions of the cooling
process yield almost identical short--time behavior and practically
identical final states of the system. The equilibration rates,
however, differ by a factor of about 1.6. It is not easy to point to
the origin of this difference. We recall that input parameters (rate
coefficients) were computed to a relative accuracy of about a few
percent. We can only speculate that the long--time behavior of the
solutions to the rate equations are sensitive to such details.

In practice, the choice between both approaches depends on the problem
under consideration. One has to solve either a large linear problem
whose dimension depends on trap cutoff and particle number, or a
smaller nonlinear one with a dimension that only depends on the trap
cutoff. The treatment of large systems requires the computation of a
large number of rate coefficients. Sums of rate coefficients enter the
rate equations in the microcanonical approach. Larger systems may
become accessible more easily once these sums can be obtained without
computing all terms individually. Work along this direction is in
progress. Likewise, we have not tested yet the validity of the
factorization assumption by solving the combined system of equations,
Eqs.~(\ref{eq35a}) and (\ref{eq35},\ref{eq36}). 

We have assumed throughout that the interaction $H_{A-A}'$ between
particles in the cooled system is weak. It is known~\cite{Stringari},
however, that no matter how small $H_{A-A}'$ is, this interaction
cannot be neglected in the condensed state once the number of
condensed atoms is sufficiently large. Does this statement -- and the
corresponding consideration for the superconducting state of Fermions
-- seriously limite our approach? We believe not. More precisely:
There exists a generalization of our approach which overcomes the
problem. It is based on the observation that it is sufficient to deal
with the system in mean--field approximation since thermodynamic
properties are unaffected by collective excitations~\cite{Stringari}.
Our rate coefficients are defined in terms of single--particle states
and single--particle energies. We may take these as solutions of
mean--field equations. Thus, we can extend our scheme as follows. We
follow the cooling process as described by our rate equations to the
point where the occupation number $n_0$ of the ground state exceeds
the critical value at which interactions become important. We define a
new set of single--particle wave functions and energies
self--consistently for that value of $n_0$ and calculate the rate
coefficients. In this way we may proceed to arbitrarily large values
of $n_0$, and similarly for a system of Fermions. Obviously,
implementation of this scheme is practical only if we succeed in
speeding up the calculation of the rate coefficients or of sums over
these coefficients.

The authors are grateful to A. Most and M. Weidem\"uller for
suggesting this problem, and for many stimulating discussions. They
also thank S. J. Wang for valuable suggestions. ANS acknowledges
support by a fellowship of FAPESP (Funda\c{c}ao de Amparo a Pesquisa
do Estado de Sao Paulo). TP thanks the Max--Planck--Institut f\"ur
Kernphysik, Heidelberg, for its hospitality during the initial stages
of this work and acknowledges support as a Wigner Fellow and staff
member at the Oak Ridge National Laboratory, managed by UT-Battelle,
LLC for the U.S. Department of Energy under Contract
DE-AC05-00OR22725.

\end{document}